\documentclass[lettersize,journal]{IEEEtran}

\usepackage{amsmath,amsfonts}
\usepackage{bm}
\usepackage{algorithm}
\usepackage{algpseudocode}
\usepackage{array}
\usepackage{textcomp}
\usepackage{stfloats}
\usepackage{url}
\usepackage{verbatim}
\usepackage{graphicx}
\usepackage{cite}
\usepackage{color}
\usepackage{xcolor}
\usepackage{eqparbox}
\usepackage{multirow}
\usepackage[numbers, sort&compress]{natbib}  
\usepackage{subcaption}
\usepackage{subfig}
\usepackage{enumitem}
\usepackage{mwe}
\usepackage{booktabs}
\usepackage{setspace}
\usepackage{siunitx}
\usepackage{fancyhdr}  

\DeclareCaptionLabelFormat{r-parens}{#2)} 
\definecolor{lightblue}{rgb}{0.0, 0.75, 0.96}
\definecolor{nicegreen}{rgb}{0.13, 0.7, 0.13}
\definecolor{darkgreen}{rgb}{ .0, .5, .0}

\usepackage{xcolor}
\usepackage{etoolbox} 

\algnewcommand{\LineComment}[1]{\State \# #1}
\algnewcommand\algorithmicswitch{\textbf{switch}}
\algnewcommand\algorithmiccase{\textbf{case}}
\algnewcommand\algorithmicassert{\texttt{assert}}
\algnewcommand\Assert[1]{\State \algorithmicassert(#1)}%
\algdef{SE}[SWITCH]{Switch}{EndSwitch}[1]{\algorithmicswitch\ #1\ \algorithmicdo}{\algorithmicend\ \algorithmicswitch}%
\algdef{SE}[CASE]{Case}{EndCase}[1]{\algorithmiccase\ #1:}{\algorithmicend\ \algorithmiccase}%
\algtext*{EndSwitch}%
\algtext*{EndCase}%

\begin{document}

\title{Diffusion-based Signal Refiner\\for Speech Enhancement and Separation}


\author{
    Masato Hirano$^*$,\ 
    Ryosuke Sawata$^*$,~\IEEEmembership{Member,~IEEE,}
    Naoki Murata,\\ 
    Shusuke Takahashi,~\IEEEmembership{Member,~IEEE,}
    Yuki Mitsufuji, ~\IEEEmembership{Senior Member,~IEEE}
\thanks{$^*$Equal contribution.}
\thanks{
Masato Hirano and Shusuke Takahashi are with Sony Group Corporation.
}
\thanks{
Ryosuke Sawata, Naoki Murata, and Yuki Mitsufuji are with Sony AI.
}
}



\maketitle

\thispagestyle{fancy}
\fancyhf{}  
\renewcommand{\headrulewidth}{0pt}  
\cfoot{
  \footnotesize
  \copyright~2026 IEEE. 
  Personal use of this material is permitted. 
  Permission from IEEE must be obtained for all other uses, in any current or future media, including reprinting/republishing this material for advertising or promotional purposes, creating new collective works, for resale or redistribution to servers or lists, or reuse of any copyrighted component of this work in other works.
}
\lhead{\footnotesize Accepted to IEEE/ACM Transactions on Audio, Speech, and Language Processing (TASLP), 2026}

\begin{abstract}
Although recent speech processing technologies have achieved significant improvements in objective metrics, there still remains a gap in human perceptual quality.
This paper proposes \emph{Diffiner}, a novel solution that utilizes the powerful generative capability of diffusion models' prior distributions to address this fundamental issue.
Diffiner leverages the probabilistic generative framework of diffusion models and learns natural prior distributions of clean speech to convert outputs from existing speech processing systems into perceptually natural high-quality audio.
In contrast to conventional deterministic approaches, our method simultaneously analyzes both the original degraded speech and the pre-processed speech to accurately identify unnatural artifacts introduced during processing.
Then, through the iterative sampling process of the diffusion model, these degraded portions are replaced with 
perceptually natural and high-quality speech segments.
Experimental results indicate that Diffiner can recover a clearer harmonic structure of speech, which is shown to result in improved perceptual quality w.r.t. several metrics as well as in a human listening test.
This highlights Diffiner's efficacy as a versatile post-processor for enhancing existing speech processing pipelines.\looseness=-1
\end{abstract}


\section{Introduction}
\label{sec:intro}
\IEEEPARstart{S}{peech} recordings in real environments contain various acoustic degradations such as background noise, reverberation, and other human speech that interferes with recording the target.
The degradations affect both the intelligibility of the target speech and various downstream tasks, e.g., speech recognition~\cite{asr_survey}, speaker identification~\cite{speaker_id_survey}, and diarization~\cite{speaker_diar_survey}.
Speech enhancement (SE) and speech separation (SS) are two fundamental acoustic signal processing tasks that remove the above acoustic degradations.
Since improvements to their performance can positively affect the downstream audio tasks mentioned above, they have been extensively studied as front-end modules~\cite{se_survey, ss_survey}.\looseness=-1

Initial studies on SE and SS explored statistical approaches such as Wiener filtering, maximum likelihood estimation, and Bayesian estimation~\cite{se_traditional_survey}.
More recently, a number of deep neural network (DNN)-based approaches have been pursued~\cite{lu2023explicit,zhao2021monaural,hu2020dccrn,stoller2018wave,choi2018phase,pascual2017segan, luo2019conv,subakan2021attention,luo2020dual,hershey2016deep}. 
Compared to the current non-DNN-based methods, these DNN-based approaches dramatically improve certain reference-based metrics, e.g., signal-to-distortion ratio (SDR)~\cite{le2019sdr}, 
perceptual evaluation of speech quality (PESQ)~\cite{rix2001perceptual}, 
and short-time objective intelligibility (STOI)~\cite{jensen2016algorithm}, that require the ground-truth signal. 
In other words, these DNN-based methods can effectively suppress the noise level and thereby improve the above metrics because, since almost all conventional DNN-based methods are based on the supervised learning, the target signals are close to the corresponding ground-truth signals, i.e., references.
However, although the level of input noise is greatly reduced, perceptually unnatural distortions originating from the DNN's non-linear signal processing tend to be included in the results.
This may adversely affect the reference-free metrics, including human listening, which do not require the corresponding ground-truth input as a reference.

To overcome the above problem, we focus on deep generative models such as generative adversarial networks (GANs)~\cite{creswell2018generative}, variational auto-encoders (VAEs)~\cite{kingma2013auto}, and diffusion models~\cite{ho2020denoising}.
This is because the degree of noise suppression through DNN-based audio processing is so strong that components which should be left in the processed speech are removed.
Namely, it is considered that removing the input noise and recovering the missed speech components may be an effective solution to the above problem.
In fact, some researchers have hypothesized that synthesizing conditioned speech would improve perceptual quality and have used a vocoder for the SE task to generatively recover clean speech~\cite{se_vocoder_1, liu2022voicefixer, lutati2023separate}.
However, building a generative model requires a huge amount of training data.
Furthermore, if a generative model-based vocoder is used as post-processing after SE/SS, the training dataset should include as many signals pre-processed by the target SE/SS model as possible.
Hence, training such deep generative model-based vocoders with the pre-processed noisy speech tends to be more laborious than in the case of using the SE/SS model only; the vocoders trained on an insufficient dataset often degrade perceptual quality.
Therefore, a new scheme, which can improve the perceptual speech quality without the laborious data collection and training, has been desired.\looseness=-1

Motivated by this background, we previously proposed a diffusion-based speech refinement technique for SE, named \textit{Diffiner}~\cite{sawata23_interspeech}.
Diffiner is based on denoising diffusion restoration model (DDRM)~\cite{kawar2022denoising}, which can provide effective diffusion updates for various degradation settings without retraining the model; thus, Diffiner can generatively improve the input signal without specialized retraining no matter what kind of preceding SE model is used.
As such, Diffiner is a versatile post-processor for any of the existing SE methods.
Our previous experimental results showed that the parts generated by Diffiner actually improve reference-free metrics related to listening perceptual quality.\looseness=-1

However, there are still some concerns that were not addressed in our prior work~\cite{sawata23_interspeech}, including (i) the effectiveness of Diffiner in not only removing background noise but also separating speech, (ii) its scalability to a large-scale data regime, and (iii) the need to confirm its validity by conducting actual human listening tests.
Hence, we address these remaining items in this paper.
Regarding (i), we found that the original formulation of Diffiner did not work well at SS.
This is because the input signal in the SS task consists of a mixture of speech from more than two individuals, and thus the assumed noise distribution does not follow the independent Gaussian originally assumed in the definition of DDRM.
Therefore, we newly extend Diffiner so that it can deal with speech signals pre-processed by not only SE but also SS.
Furthermore, to resolve (ii), we retrain Diffiner by using a larger dataset than the original one and assess its validity in experiments.
To address (iii), we conduct not only quantitative evaluations but also a subjective listening test.

In Sec.~\ref{sec:related} of this paper, we provide a brief review of related work, with a particular focus on the differences between Diffiner and recent diffusion-based SE and SS methods.
Section~\ref{sec:background} explains the core components of our method, i.e., diffusion-based generative models and DDRM. 
Section~\ref{sec:proposal} derives Diffiner for both SE and SS.
In Sec.~\ref{sec:exp}, we demonstrate the effectiveness of Diffiner through qualitative, quantitative, and human listening evaluations.
We conclude in Sec.~\ref{sec:conclusion} with a summary and mention of future work.

\section{Related work}
\label{sec:related}
For the sake of the following discussion, we first clarify the difference between SE and SS.
SE aims to suppress \emph{non-speech noise}, while SS targets the extraction of the desired speech from \emph{mixtures composed of multiple speakers}.

\subsection{Non-DNN-based methods}
\label{subsec:non_dnn}
Before DNN-based methods appeared, researchers studied classical signal processing methods for SE and SS~\cite{se_traditional_survey, wiener_org, subtraction_org, nmf_org}.
Almost all of them have been already replaced with DNN-based SE and SS methods, as described in the subsequent section.
However, it is worth noting that one of them, \emph{Wiener filter}~\cite{wiener_org}, is strongly related to Diffiner in that the optimal coefficients at each step of the reverse process in Diffiner are equivalent to those of Wiener filter.
Specifically, Diffiner filters the degraded speech signals so that the parts highly related to the target speech remain and the non-related parts are replaced with plausible ones generated by Diffiner.
The processing by Diffiner can thus be regarded as something like a \emph{generative Wiener filter}, as the vanilla Wiener filter was also designed so that the target parts remain and the non-related parts are suppressed based on the prior signal-to-noise ratio (SNR).

Therefore, in contrast to Wiener filter, Diffiner can not only suppress non-reliable parts but also replace them with high-fidelity speech components based on prior information.
More details regarding Diffiner will be described in Sec.~\ref{sec:proposal}.

\subsection{DNN-based methods}
\label{subsec:dnn}
The DNN-based methods appearing in recent years have attempted to overcome the above limitation of classical SS and boost the performance of SE and SS.

In the field of SE, simply introducing a DNN significantly improved the degree of noise suppression.
For example, Luo et al. showed that utilizing the learnable convolution filters of convolutional neural networks (CNNs) can achieve better encoders and decoders than those that use short-time Fourier transform (STFT) and inverse STFT (ISTFT)~\cite{luo2019conv}.
Furthermore, it has been reported that the architecture of recurrent neural networks (RNNs) is suitable for real-time DNN-based speech processing \cite{rnn_se_1, rnn_se_2, rnn_se_3, crnn_hybrid_se_3}.\looseness=-1

In the field of SS, Yu et al. proposed permutation invariant training (PIT) for the task of source separation~\cite{pit_org}, and Hershey et al. proposed deep clustering (DC)~\cite{hershey2016deep}.
PIT considers all possible correspondences between the output of a DNN and the speakers included in the input mixture, whereas DC estimates embedded features such that they are grouped into the same class if they come from the same person.

In these ways, DNN-based SE and SS have exhibited dramatically improved performances over conventional methods.

\subsection{Deep generative model-based methods}
\label{subsec:generative}
As mentioned in Sec.~\ref{sec:intro}, recent DNN-based methods have outperformed classical non-DNN-based ones by sufficiently suppressing the noise mixed in the input.
However, such methods often remove parts of the target speech when the spectrogram contains overlapping noise.
The results sound clear but may seem unnatural due to distortion caused by a lack of necessary components in the target spectrogram.

To deal with this issue, some approaches have tried to use deep generative models such as GAN and VAE to recover the missing components of human speech~\cite{se_vocoder_1, liu2022voicefixer}.
In this context, for the purpose of speech refinement, Shi et al. and Liu et al. utilized deep generative model-based vocoders, i.e., GAN or VAE, to recover the human speech erased by the preceding SE or SS models.\looseness=-1

More recently, researchers have focused on utilizing diffusion models for generative SE and SS.
As far as we know, there are roughly two ways to do this: i) a unified SE or SS using a conditioned diffusion-based model~\cite{serra2022universal, richter2023speech, scheibler2022diffusion}, and ii) two-stage processing involving enhancement (or separation) and a subsequent diffusion-based vocoder~\cite{lemercier2022storm, lutati2023separate}.
Regarding i), universal SE with score-based diffusion (UNIVERSE)~\cite{serra2022universal}, score-based generative model for SE (SGMSE)~\cite{richter2023speech}, and diffusion-based generative speech source separation (DiffSep)~\cite{scheibler2022diffusion} have been proposed to conduct conditioned generative SE or SS.
Regarding ii), diffusion-based stochastic regeneration model (StoRM)~\cite{lemercier2022storm} has been proposed for two-stage SE with a post-diffusion vocoder.
Lutati et al. proposed a framework that effectively uses a pretrained speech diffusion model as a post-refiner on the speech separation results~\cite{lutati2023separate}.

Other researchers have tried to accelerate diffusion-based SE and SS~\cite{soundctm, sb_se, sbctm}.
The reasoning is that diffusion-based SE and SS often have much higher inference computational costs than other SE and SS models---although they have improved the conventional performance dramatically, as shown above---and this might be disadvantageous in cases of a downstream task that requires real-time responses, e.g., tele-communication and automatic speech recognition (ASR) systems. 
In the field of SE and SS, there are two main methods to accelerate diffusion-based models: decreasing the number of calculation iterations~\cite{li2024diffusion, soundctm} and finding an optimal alternative path instead of the forward/reverse process~\cite{sb_se, sbctm}.
Consistency trajectory models (CTMs)~\cite{ctm} enable a diffusion-based model to jump between arbitrary time steps in the reverse process, and thus a model based on CTM can ideally obtain output in just one step.
By taking advantage of this feature of CTM, SoundCTM~\cite{soundctm} achieved almost the same level of performance as the original model on a text-to-sound task, even though its calculation was performed in just one step.
Furthermore, Li et al. decreased the multiple iterations of the forward/reverse process by instead utilizing the score derived from a discriminative model~\cite{li2024diffusion}.
On the other hand, Schr\"{o}dinger bridge (SB)~\cite{schrodinger_bridge} is a method to seek another optimal path between probability distributions.
Juki\'{c} et al. proposed a generative SE model based on SB, called SB-based SE~\cite{sb_se}, and Nishigori et al. used a hybrid of CTM and SB for the SE task to create Schr\"{o}dinger bridge consistency trajectory models (SBCTM)~\cite{sbctm}.
SBCTM achieved a roughly $16\times$ improvement in RTF and allowed a controllable trade-off between quality and calculation cost by adjusting the number of steps.\looseness=-1

In addition, there is another concern w.r.t. data collection when training the conditioned model.
Specifically, building a unified SE or SS model based on a deep generative model conditioned by noisy speech tends to require a huge amount of both noisy and clean speech compared to training a conventional DNN-based model~\cite{se_datasize}.
This often makes developing a high-performance model difficult, and the perceptual quality may be degraded if the dataset used to train the model is not large enough.
On the other hand, two-stage processing, i.e., performing conventional SE or SS first and generative refinement second, does not spoil the performance of the first stage because these processes are independent, and thus we can adaptively decide to use the post-refiner depending on the downstream task, i.e., the tradeoff between performance, inference speed, and data collection cost.
Hence all one has to do is build a versatile and effective second stage, i.e., a generative model-based speech refiner.

In fact, some researchers have proposed such diffusion-based speech refiners \cite{se_vocoder_1, liu2022voicefixer, lutati2023separate, lemercier2022storm}.
For example, Lemercier et al. proposed an SE method that uses a \textit{stochastic regeneration} approach, StoRM~\cite{lemercier2022storm}, in which the diffusion model's forward and backward processes are applied to the output of the model preceding it.
However, since the preceding model and the subsequent diffusion-based refiner in StoRM are simultaneously trained in an end-to-end (E2E) manner, StoRM needs to retrain the diffusion model whenever the preceding model is changed.
For the same reason, its refiner cannot be applied to any other method preceding it except the one used for E2E training.
Lutati et al.~\cite{lutati2023separate} proposed \emph{DiffWave}~\cite{kong2020diffwave} which is a two-stage method combining an SS method with a diffusion-based post-speech refiner.
The authors exploited the versatility of DiffWave and showed that it improved the performance scores of some of the SS methods preceding the DiffWave-based post-refiner.
Their approach is similar to our own idea of a two-stage speech processing consisting of a preceding speech processor and \emph{Diffiner}.
However, the overall performance of their method tended to be affected by leftover noise from the preceding speech processor since the post speech processor, i.e., DiffWave, was trained on only clean speech, and it assumes that there is no noise in the signal to be fed into it.
Namely, the leftover noise from the preceding speech processor degrades the performance of DiffWave.
Therefore, the aforementioned refiners need to be retrained depending on the type of noise left over from the preceding SE or SS module.
This means that the aforementioned tendency of deep generative models requiring a huge amount of training data remains unresolved.
\\

To overcome this problem, we previously proposed a DDRM-based speech refiner, Diffiner~\cite{sawata23_interspeech}, for SE.
Once Diffiner is trained on only clean speech, it can be applied to any existing SE without specialized retraining.
This is because, unlike DiffWave, the DDRM-based speech refiner can adaptively consider the distribution of the leftover noise mixed from the preceding method.
To confirm this, we briefly conducted a preliminary experiment comparing DiffWave and DDRM used in Diffiner by inputting many signals that have diverse signal-to-noise ratios (SNRs).
\begin{table}[t]
\caption{
    NISQA results at different Gaussian noise levels.
    Higher values are better.
    DiffWave and DDRM were trained on LJSpeech~\cite{ljspeech} which consists of only clean speech. 
}
 \label{table:preliminary_experiment}
 \centering
 \small
 \begin{tabular*}{\columnwidth}{@{\extracolsep{\fill}}c|ccc}
 \toprule
 SNR [dB] & Source & DiffWave \cite{lutati2023separate} & DDRM in Diffiner (ours) \\
 \midrule
 $\infty$ (clean) & 4.379 & 4.374 & \textbf{4.386}\\
 20 & 2.915 & 3.036 & \textbf{4.117}\\
 15 & 2.295 & 2.736 & \textbf{3.995}\\
 10 & 1.761 & 2.091 & \textbf{3.832}\\
 5 & 1.335 & 1.693 & \textbf{3.612}\\
 \bottomrule
 \end{tabular*}
\end{table}
The results are listed in Table~\ref{table:preliminary_experiment} (note that the NISQA metric is described in Sec.~\ref{sec:exp}).
As shown here, the scores of DiffWave decreased as the SNR decreased, whereas the DDRM used in Diffiner maintained good scores regardless of the SNR value.
We therefore incorporated the DDRM-based speech refiner in Diffiner and confirmed that it works well no matter what kind of SE model is used \cite{sawata23_interspeech}.
However, its validity was confirmed only on SE, and thus we newly propose a further extension of Diffiner for both SE and SS in the current work.\looseness=-1

Compared to the related methods mentioned above, the contributions of Diffiner are threefold:
a) training using clean speech only (i.e., noisy speeches are unnecessary),
b) high modularity, where Diffiner can be detached and then applied to any other existing SE and SS models as a post-refiner,
and c) preceding model-agnostic versatility, wherein Diffiner does not require specialized retraining regardless of what model of SE or SS is used (i.e., Diffiner trained on only clean speech can refine the results of any SE and SS methods preceding Diffiner).\looseness=-1

\section{Background}
\label{sec:background}
Let us revisit the mathematical formulation of DDPM~\cite{ho2020denoising} and DDRM~\cite{kawar2022denoising}.
DDRM is a method for conditional generation based on diffusion models. 
It leverages a pre-trained DDPM as a prior distribution and generates data that serves as a solution to a linear inverse problem defined by a linear degradation model. 
Below, we review the linear inverse problem and the DDPM.
Then, we explain the conditional generation using the DDRM.\looseness=-1

\subsection{Linear inverse problem}
\label{subsec:linear_inverse_problem}
The linear inverse problem involves estimating an unknown signal $\bm{x}_0\in\mathbb{R}^{d_x}$ given conditioning information $\bm{y}\in\mathbb{R}^{d_y}$. The relationship between $\bm{x}_0$ and $\bm{y}$ is defined by the following degradation model:
\begin{align}
 \bm{y} = \bm{Hx}_0 + \bm{z},
 \label{eq:lip}
\end{align}
where $\bm{H}\in\mathbb{R}^{d_y\times d_x}$ is a known degradation matrix, 
and $\bm{z}\sim {\cal N}(\bm{0}, \sigma_{\bm{y}}^2\bm{I})$ is 
\textit{i.i.d.} additive Gaussian noise of known variance $\sigma_{\bm{y}}^2$.\looseness=-1

\subsection{Learning data distribution by using DDPM}
\label{subsec:forward_reverse}
DDPM~\cite{ho2020denoising} is a deep generative model for learning the model distributions 
$p_\theta(\bm{x})$ of data $\bm{x}$, which approximate an unknown data distribution $q(\bm{x})$. In the DDPM, data are produced in a Markov chain manner, starting from Gaussian noise followed by iterative denoising into the desired data. 
Let $\bm{x}_0$ be the target sample and $\bm{x}_T\rightarrow\bm{x}_{T-1}\rightarrow\cdots\rightarrow\bm{x}_0$ be the backward process that progressively denoises toward $\bm{x}_0$.
The model distribution $p_\theta(\bm{x}_0)$ is defined as:
\begin{align}
 &p_\theta(\bm{x}_0) = \int p_\theta(\bm{x}_{0:T})\mathrm{d}\bm{x}_{1:T}\nonumber\\
 &\text{where}\ p_\theta(\bm{x}_{0:T})\equiv p_\theta(\bm{x}_T)\prod_{t=0}^{T-1}p_\theta(\bm{x}_t|\bm{x}_{t+1}).
\label{eq:reverse_process_joint_prob}
\end{align}
The parameters $\theta$ are learned by optimizing a variational lower bound with inference distribution $q(\bm{x}_{1:T}|\bm{x}_0)$, which is decomposed into a Markov chain of Gaussian transitions:
\begin{align} 
 &q(\bm{x}_{1:T}|\bm{x}_0) \equiv \prod_{t=0}^{T-1} q(\bm{x}_{t+1}|\bm{x}_{t}), \nonumber\\
  \text{where}\ &q(\bm{x}_{t+1}|\bm{x}_{t}) \equiv 
  {\cal N} \left(\bm{x}_t,\left(\sigma^2_{t+1}-\sigma^2_{t}\right)\bm{I} \right),
 \label{eq:forward_process}
\end{align}
where $\sigma_1,\sigma_2,\ldots,\sigma_T\in(0,\infty]$ is an increasing parameter. This is called a \textit{forward process}, which is typically fixed during training, while $p_\theta(\bm{x}_{0:T})$ is a \textit{generative process} that approximates the \textit{reverse process} $q(\bm{x}_{t}|\bm{x}_{t+1})$.
By marginalizing the forward process, we obtain $q(\bm{x}_t|\bm{x}_0) = {\cal N}(\bm{x}_t;\ \bm{x}_0,\sigma^2_t\bm{I})$, which allows sampling as
\begin{align}
    \bm{x}_t &= \bm{x}_0 + \sigma_t\epsilon_t 
    \:\: \mbox{s.t.} \:\: \epsilon_t \sim {\cal N}(\bm{0},\bm{I}).
\end{align}
In practice, modeling all conditionals as Gaussian distributions leads to the training objective:
\begin{align}\label{eq:elbo} 
 \sum_{t=1}^T\gamma_t\mathbb{E}_{(\bm{x}_0,\bm{x}_t)\sim q(\bm{x}_0)q(\bm{x}_t|\bm{x}_0)}
 \left[ \|\bm{x}_0 - f_\theta(\bm{x}_t,t)\|_2^2 \right],
\end{align}
where $f_\theta$ is a function for estimating $\bm{x}_0$ and $\gamma_{1:T}$ are positive coefficients depending on the forward process $q(\bm{x}_{1:T}|\bm{x}_0)$.




\subsection{Conditional generation with DDRM}
\label{subsec:conditional_ddrm}
The DDRM~\cite{kawar2022denoising} is a diffusion-based conditional sampling method with a Markov chain structure conditioned on 
$\bm{y}$, where the joint distribution is defined as 
\begin{align}
 p_\theta(\bm{x}_{0:T}|\bm{y}) = p_\theta(\bm{x}_T|\bm{y})\prod_{t=0}^{T-1}p_\theta(\bm{x}_{t}|\bm{x}_{t+1}, \bm{y}).
 \label{eq:ddrm_joint}
\end{align}
The likelihood $p(\bm{y}|\bm{x}_0)$ is characterized as a linear inverse problem with a known degradation matrix $\bm{H}$.
Here, we consider a variational distribution conditioned on $\bm{y}$ as 
\begin{align}
    q(\bm{x}_{1:T}|\bm{x}_0,\bm{y}) = q(\bm{x}_T|\bm{x}_0,\bm{y})
    \prod_{t=0}^{T-1}q(\bm{x}_t|\bm{x}_{t+1},\bm{x}_0,\bm{y}),
\end{align}
where the corresponding variational lower bound is defined as
\begin{align}
    &\max_{\theta} 
      \mathbb{E}_{q(\bm{x}_0),\,q(\bm{y}|\bm{x}_0)}
      [\log p_{\theta}(\bm{x}_0|\bm{y})]
      \nonumber \\
    &\geq
      \max_{\theta}
      \mathbb{E}_{\substack{
          q(\bm{x}_0),\,
          q(\bm{y}|\bm{x}_0),\\
          q(\bm{x}_{1:T}|\bm{x}_0,\bm{y})
      }}
      \big[
      \log p_{\theta}(\bm{x}_{0:T}|\bm{y})
      - \log q(\bm{x}_{1:T}|\bm{x}_0,\bm{y})
      \big].
\end{align}
The key point of the DDRM is that, given the above conditional forward process, 
the variational distribution $q$ is constructed such that $q(\bm{x}_t|\bm{x}_0) = {\cal N}(\bm{x}_0, \sigma_t^2\bm{I})$ when the process is marginalized over $\bm{x}_{t'}$, for all $t'>t$ and $\bm{y}$.
The marginalized distribution has the same form as in the DDPM; thus, in the DDRM, we can use the pretrained DDPM for any linear inverse problem with a theoretical guarantee of it having the same variational lower bound as the DDPM (see the derivations in~\cite{kawar2022denoising}).\looseness=-1



The DDRM is updated in the spectral space of 
the singular value decomposition (SVD) of $\bm{H} = \bm{U\Sigma V}^\top$, 
where $\bm{U}\in\mathbb{R}^{d_U\times d_U},\ \bm{V}\in\mathbb{R}^{d_V\times d_V}$ are 
orthogonal matrices and $\bm{\Sigma}\in\mathbb{R}^{d_U\times d_V}$ is a rectangular matrix containing the singular values $s_1 \geq s_2 \geq \cdots \geq s_{d_V}$ of $\bm{H}$ on its main diagonal.
To simplify the notation in the spectral space, let $\bar{\bm{x}}_t=\bm{V}^\top\bm{x}_t$, $\bar{\bm{y}}=\bm{\Sigma}^\dag\bm{U}^\top\bm{y}$ and $s_i$ be the $i$th singular value of $\bm{H}$, where $\bm{A}^\dag$ is the Moore-Penrose pseudo-inverse matrix of $\bm{A}$. 
The DDRM treats each element in the spectral space differently depending on its corresponding singular value.
When $s_i = 0$, the measurements $\bm{y}$ contain no information about the $i$th element, 
so the update follows the standard unconditional diffusion generation. 
When $s_i > 0$, the measurements provide partial information, 
and the update is determined by comparing the observation noise level $\sigma_y/s_i$ 
with the diffusion noise level $\sigma_t$.\looseness=-1

We denote the $i$th element of a vector $\bm{a}$ as $a^{(i)}$; accordingly, the DDRM defines the variational distribution as
\begin{align}
    q(\bar{x}_T^{(i)}|\bm{x}_0,\bm{y}) =
    &\begin{cases}
        {\cal N}(\bar{y}^{(i)},\sigma^2_T-\frac{\sigma_{\bm{y}}^2}{s_i^2})\ &\text{if}\ s_i>0\\
        {\cal N}(\bar{x}^{(i)}_0,\sigma_T^2)\ &\text{if}\ s_i=0
    \end{cases},
\end{align}
\begin{align}
    &q(\bar{x}_t^{(i)}|\bm{x}_{t+1},\bm{x}_0,\bm{y}) = \nonumber\\
    &\begin{cases}
    {\cal N}(\bar{x}_0^{(i)} + \sqrt{1-\eta^2}\sigma_t\dfrac{\bar{x}_{t+1}^{(i)}-\bar{x}_0^{(i)}}{\sigma_{t+1}}, \eta^2\sigma_t^2)\ &\text{if}\ s_i=0\\
    {\cal N}(\bar{x}_0^{(i)} + \sqrt{1-\eta^2}\sigma_t\dfrac{\bar{y}^{(i)}-\bar{x}_0^{(i)}}{\sigma_{\bm{y}}/s_i}, \eta^2\sigma_t^2)\ &\text{if}\ \sigma_t<\frac{\sigma_{\bm{y}}}{s_i}\\
    {\cal N}((1-\eta_b)\bar{x}_0^{(i)} + \eta_b\bar{y}^{(i)}, \sigma_t^2-\frac{\sigma_{\bm{y}}^2}{s_i^2}\eta_b^2)\ &\text{if}\ \sigma_t\geq\frac{\sigma_{\bm{y}}}{s_i}
    \end{cases},
\end{align}
where $\eta$ and $\eta_b$ are hyperparameters ranging in $(0,1]$.
The conditional distributions defined above have been proven to satisfy $q(\bm{x}_t|\bm{x}_0) = {\cal N}(\bm{x}_0, \sigma_t^2\bm{I})$ (see~\cite{kawar2022denoising}) as is the case with the DDPM.  
Let $\bm{x}_{\theta,t}$ be the $t$th prediction of the pretrained DDPM $f_{\theta}(\bm{x}_{t+1},t+1)$, and $\bar{\bm{x}}_{\theta,t}=\bm{V}^\top\bm{x}_{\theta,t}$ be the projection of $\bm{x}_{\theta,t}$ to the spectral space. 
DDRM then defines the model distribution $p_\theta$ with a trainable parameters $\theta$ as follows:
\begin{align}
    p_{\theta}(\bar{x}_T^{(i)}|\bm{y}) =
    &\begin{cases}
        {\cal N}(\bar{y}^{(i)},\sigma^2_T-\frac{\sigma_{\bm{y}}^2}{s_i^2})\ &\text{if}\ s_i>0\\
        {\cal N}(0,\sigma_T^2)\ &\text{if}\ s_i=0
    \end{cases},
\end{align}
\begin{align}\label{eq:ddrm_modeldist}
    &p_{\theta}(\bar{x}_t^{(i)}|\bm{x}_{t+1},\bm{y}) = \nonumber\\
    &\begin{cases}
    {\cal N}(\bar{x}_{\theta,t}^{(i)} + \sqrt{1-\eta^2}\sigma_t\dfrac{\bar{x}_{t+1}^{(i)}-\bar{x}_{\theta,t}^{(i)}}{\sigma_{t+1}}, \eta^2\sigma_t^2)\ &\text{if}\ s_i=0\\
    {\cal N}(\bar{x}_{\theta,t}^{(i)} + \sqrt{1-\eta^2}\sigma_t\dfrac{\bar{y}^{(i)}-\bar{x}_{\theta,t}^{(i)}}{\sigma_{\bm{y}}/s_i}, \eta^2\sigma_t^2)\ &\text{if}\ \sigma_t<\frac{\sigma_{\bm{y}}}{s_i}\\
    {\cal N}((1-\eta_b)\bar{x}_{\theta,t}^{(i)} + \eta_b\bar{y}^{(i)}, \sigma_t^2-\frac{\sigma_{\bm{y}}^2}{s_i^2}\eta_b^2)\ &\text{if}\ \sigma_t\geq\frac{\sigma_{\bm{y}}}{s_i}
    \end{cases}.
\end{align}


\section{Diffiner: DDRM-based speech refiner}
\label{sec:proposal}
This section presents \emph{Diffiner}, our novel DDRM-based speech refiner.
Once Diffiner is trained on only clean speech, it can be applied without task-specified training to any signal results that have been pre-processed by either SE or SS.
This is because all Diffiner has to do is learn the prior distribution of clean speech, regardless of the kind of preceding speech processor, for it to be utilized to restore any degraded signals.

To use Diffiner for both SE and SS, it is necessary to make a slight modification only to the inference step.
Therefore, we first present the Diffiner algorithm for SE, and then derive one for SS by redesigning the observation model.

\subsection{Preliminaries}
\label{subsec:notations}
Before explaining Diffiner, we summarize the notations used throughout the theoretical discussion related to SE and SS.
We denote time-domain speech signals with lowercase letters, e.g., $\bm{a}\in\mathbb{R}^{d_a}$, where $d_a$ is the length of the signal.
We can use any SE and SS model for conditioning the DDRM update in Diffiner. 
These preceding models estimate the clean speech signal given noisy speech. 
We distinguish the estimated speech from the corresponding speech signal by adding $\hat{\cdot}$ to the letters, e.g., $\hat{\bm{a}}\in\mathbb{R}^{d_a}$.
Our refiner operates in the time-frequency (TF) domain, where each speech signal is transformed into STFT coefficients. We use uppercase letters, e.g., $\bm{A}\in\mathbb{C}^{d_k\times d_l}$ to represent the STFT coefficients of the corresponding time-domain signals $\bm{a}$, where $d_k$ and $d_l$ are the numbers of frequency bins and time frames, respectively. 
In the design of a linear inverse problem for the DDRM, we assume that an observation of one STFT coefficient is independent of the other coefficients. We utilize $k=1,\ldots,d_k$ and $l=1,\ldots,d_l$ to denote the $(k,l)$th STFT coefficient as $A^{(k,l)}$ and define the linear inverse problem in a TF-bin-wise manner.
We also use a non-bold uppercase letter with a suffix $t$ as $A_t$ to denote the $(k,l)$th bin of the STFT coefficient $\bm{A}$ at the diffusion time $t$, where we drop the $(k,l)$ indices for notational brevity.

For the SE and SS tasks, we first train a DDPM on clean speech data to obtain the model distribution $p_\theta$. 
Here, the forward process for $\bm{A}_t$ is defined as
\begin{align} 
 &q(\bm{A}_{1:T}|\bm{A}_0) \equiv \prod_{t=0}^{T-1} q(\bm{A}_{t+1}|\bm{A}_{t}), \nonumber\\
  \text{where}\ &q(\bm{A}_{t+1}|\bm{A}_{t}) \equiv 
  {\cal N}_{\mathbb{C}} \left(\bm{A}_t,\left(\sigma^2_{t+1}-\sigma^2_{t}\right)\bm{I} \right).
\end{align}
Accordingly, the marginal distribution becomes a circular-symmetric complex Gaussian:
\begin{align}
    q(\bm{A}_t|\bm{A}_0) = {\cal N}_{\mathbb{C}}(\bm{A}_0,\sigma_t^2\bm{I}),
\end{align}
where $\sigma_t$ is the variance representing the diffusion noise level, and we assume that $0=\sigma_0 < \sigma_1 < \cdots < \sigma_T$. On the basis of the variational distribution $q$, we first train the denoiser $f_\theta$ in the DDPM by using the loss function in Eq.~(\ref{eq:elbo}). The learned denoiser $f_\theta$ is then utilized in the DDRM update in accordance with Eq.~(\ref{eq:ddrm_modeldist}) to obtain the refined speech signal.

\subsection{Refiner for speech enhancement}
\label{subsec:refiner_enh}

Now let us define the degradation model for SE in the DDRM update.
We assume that the noisy input signal is the sum of the clean speech and noise signal, $Y^{(k,l)}=X^{(k,l)}+N^{(k,l)}$, where $Y^{(k,l)}\in\mathbb{C}$, $X^{(k,l)}\in\mathbb{C}$, and $N^{(k,l)}\in\mathbb{C}$ are the noisy input, clean speech and noise signal, respectively. Hereafter, the indices of $k$ and $l$ will be dropped for notational brevity. We must consider how to connect this noisy observation model with the degradation model defined in Eq.~(\ref{eq:lip}).
To this end, we propose using the preceding enhancement model for estimating the noise variance of $N$.
The noise can be estimated from its output $\hat{X}$ as $\hat{N}=Y-\hat{X}$.
Then, we model the estimated variance of the STFT coefficients $N$ as 
\begin{align}\label{eq:est_sigma}
    \hat{\sigma}^2 = \min(\max(\lambda |\hat{N}|^2,\delta),R),
\end{align}
where $\delta >0$ and $R>0$ are respectively the minimum and maximum thresholds of the estimated variance in order to avoid numerical instability.
Furthermore, $\lambda>0$ is a tunable hyperparameter.
The overall linear inverse problem for SE is defined as\looseness=-1
\begin{align}
    Y=X+N\ \mathrm{s.t}\ N\sim{\cal N}_{\mathbb{C}}(0, \hat{\sigma}^2).
    \label{eq:se_def}
\end{align}
After dividing both sides by $\hat{\sigma}$, we obtain
\begin{align}
\label{eq:deg_enh}
    \tilde{Y} = \dfrac{1}{\hat{\sigma}}X+Z\ \mathrm{s.t}\ Z\sim {\cal N}_{\mathbb C}(0,1),
\end{align}
where $\tilde{Y}=Y/\hat{\sigma}$. 
We assume that $Z$ follows an \textit{i.i.d.} complex Gaussian distribution for tractability.
Given this degradation model, the proposed DDRM-based refiner for SE is summarized in Algorithm~\ref{alg:proposal_alg} (see ``case \texttt{se}'').
Namely, Eqs.~(\ref{eq:se_def}) and (\ref{eq:deg_enh}) are computed for each TF bin, i.e., ($k$, $l$), and are applied to all TF bins.
The algorithm shows that when the diffusion noise is smaller than the observation noise ($\sigma_t < \hat{\sigma}$), the generative process gives more weight to the generation result $\bm{x}_{\theta,t}$ than to the conditioning information $\bm{y}$ and it considers the conditioning information to be more important when the diffusion noise is larger than the observation noise ($\sigma_t \geq \hat{\sigma}$). 
Intuitively, a small amount of observation noise means that the conditioning information is clean enough to be preserved in the generated result, whereas a large amount means that the information is no longer informative and the generated result should be given more weight.

In practical applications of DDRM updates, the observation noise vector may have correlations across different STFT coefficients. 
Consequently, the term $(X-X_{\theta,t})$ in ``\textsc{Sampler}($\cdot$)'' of Algorithm~\ref{alg:proposal_alg}, which utilizes the information from the observations, may not necessarily follow the independent complex Gaussian with the estimated variance assumed by DDRM. 
To address this issue, we propose a modification to the original DDRM framework: we use $X=X_{t+1}$ instead of $X=Y$ 
if $\sigma_t<\sigma$, as the observed signals are no longer helpful under this condition. These changes to the terms mean that the algorithm ignores the observational information and relies on the generative model during less noisy steps ($\sigma_t<\sigma$). To distinguish the modified version of the refinement algorithm from the original Diffiner, we add ``+" at the end of the term, i.e., ``Diffiner+". The algorithm of Diffiner+ is provided as ``Pattern 2" in Algorithm~\ref{alg:proposal_alg}.


\begin{figure}[!t]
\begin{algorithm}[H]
 \caption{Diffiner}
 \label{alg:proposal_alg}
 \begin{algorithmic}[1]
 \Require output(s) of preceding speech enhancement (\texttt{se}) or speech separation (\texttt{ss}) model
 \Ensure refined estimate $\bm{X}_{\textrm{ref}}$
 \State Choose which method ${\cal M}\in\{\texttt{se},\texttt{ss}\}$ to apply
 \LineComment{From here, we omit $k$ and $l$ for notational brevity}
 \Switch {${\cal M}$}
 \Case {\texttt{se}}
    \State Let the output of the preceding model be $\hat{\bm{X}}$
    \State Calculate $\hat{\sigma}^2$ in Eq.~\eqref{eq:est_sigma} by using $\hat{\bm{X}}$
    \State Initialize $X_T \sim {\cal N}_{\mathbb{C}}(0,\sigma^2_T-\hat{\sigma}^2)$ 
    \For{$t=T-1$ to $0$}
    \State Predict denoised signal $X_{\theta,t} = f_\theta(X_{t+1},t+1)$ 
    \State $\triangleright$ \textit{\textbf{Pattern 1: Diffiner}} 
    \State \hspace{\algorithmicindent} $X_t = \textsc{Sampler}(X_{\theta,t}, Y,\hat{\sigma},\sigma_t)$
    \State $\triangleright$ \textit{\textbf{Pattern 2: Diffiner+}} 
    \State \hspace{\algorithmicindent} $X_t = \textsc{Sampler}(X_{\theta,t}, X_{t+1},\hat{\sigma},\sigma_t)$ 
    \EndFor
    \State Refined signals $\bm{X}_{\textrm{ref}} = \bm{X}_{0}$
 \EndCase
 \Case {\texttt{ss}}
    \State Let preceding outputs be
    $\hat{\bm{X}}^{(1)},\ldots,\hat{\bm{X}}^{(M)}$ 
    \State Project observation as $\bar{\bm{\Psi}} = \bm{\Sigma}^\dag\bm{U}^\top\bm{\Psi}$
    \LineComment{$\bar{X}^{(i)}$ corresponds to $i$th singular value $s_i$}
    \State Initialize $\bar{X}^{(i)}_T \sim {\cal N}_{\mathbb{C}}(0,\sigma^2_T-\sigma^{(i)2}/s^2_i)$
    \For{$t=T-1$ to $0$}
    \State $\bm{X}_{t+1} \leftarrow \bm{V}\bar{\bm{X}}_{t+1}$
    \State $X_{\theta,t} \leftarrow f_\theta(X_{t+1},t+1)$
    \State $\bar{\bm{X}}_{\theta,t} \leftarrow \bm{V}^\top\bm{X}_{\theta,t}$
    \State \hspace{\algorithmicindent} $\bar{X}_t^{(i)} = \textsc{sampler}(\bar{X}^{(i)}_{\theta,t}, \bar{X}_{t+1}^{(i)},\sigma^{(i)}/s_i,\sigma_t)$
    \EndFor
    \State Refined signals $\bm{X}_{\textrm{ref}} = \bm{X}_{0} = \bm{V}\bar{\bm{X}}_0$
 \EndCase
 \EndSwitch
 \Return $\bm{X}_{\textrm{ref}}$
 \Statex
 \hrulefill
 \Function {sampler}{$X_{\theta,t}, X, \sigma, \sigma_t$}
 \If{$\sigma_t < \sigma$} 
 \State $X_t \sim {\cal N}_{\mathbb{C}}(X_{\theta,t} + \sqrt{1-\eta_a^2}\sigma_t\dfrac{X-X_{\theta,t}}{\sigma},\eta_a^2\sigma^2_t)$
 \ElsIf{$\sigma_t \geq \sigma$} 
 \State $X_t \sim {\cal N}_{\mathbb{C}}((1-\eta_b)X_{\theta,t} + \eta_b X,\sigma^2_t-\eta^2_b\sigma^2)$
 \EndIf
 \EndFunction
 \end{algorithmic}
\end{algorithm}
\end{figure}

\subsection{Refiner for speech separation}
\label{subsec:refiner_sep}
\begin{figure}[t]
 \centering
 \includegraphics[width=\columnwidth]{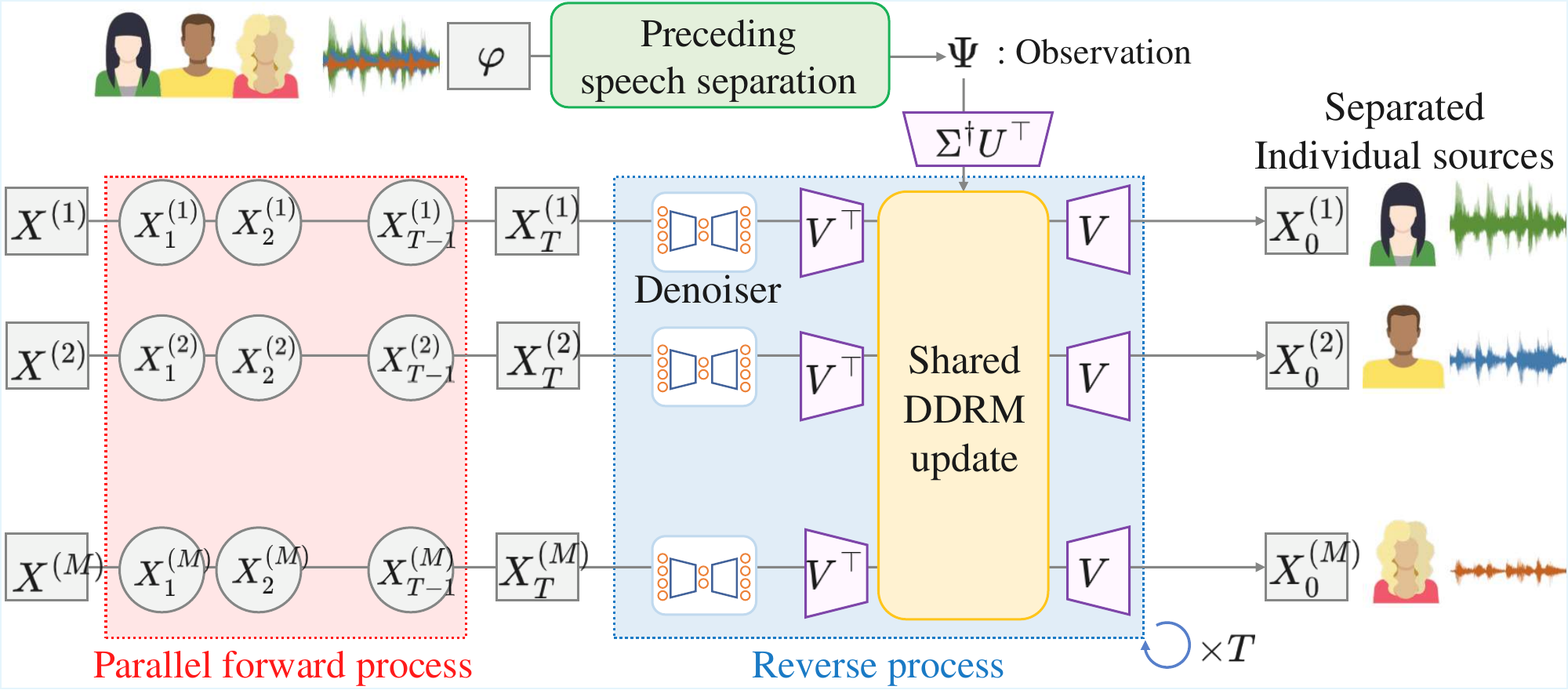}
 \caption{
    Outline of Diffiner for speech separation. 
    $M$ tracks are perturbed in parallel with Gaussian noise in the forward process (left). 
    The learned denoiser then iteratively generates the sample in the reverse process (right). 
    The outputs of the preceding separation are used for conditioning (upper).
    In this way, the generation of each source is integrally guided over the shared DDRM update.\looseness=-1}
 \label{fig:overview}
\end{figure}




In this section, we modify Diffiner so that it can be used in the SS task by redesigning the observation model.
In contrast to the case of SE, ``noise'' in SS may also be human speech, i.e., interfering speech.
Its distribution is correlated with the target speech, and thus it does not follow the independent Gaussian distribution assumed in the DDRM definition. 
Consequently, the degradation model in Eq.~(\ref{eq:deg_enh}) is no longer appropriate for this scenario.
Assuming a mixture of $M$ speakers, we have\looseness=-1
\begin{align}
    \bm{\varphi} = \sum_{j=1}^M \bm{x}^{(j)},
    \label{eq:prop_lip}
\end{align}
where $\bm{\varphi}$ and $\bm{x}^{(j)} \: (j\in\{1,\ldots,M\})$ are the mixture and clean speech of the $j$th speaker, respectively.
We consider the STFT coefficients of both $\bm{\varphi}$ and $\bm{x}^{(j)}$, which are denoted as $\bm{\Phi}\in\mathbb{C}^{d_k\times d_l}$ and $\bm{X}^{(j)}\in\mathbb{C}^{d_k\times d_l}$, respectively.
Given estimates of the individual speech $\hat{\bm{X}}^{(j)}$ from the $j$th speaker, 
we can use a simple degradation model in which each estimate is observed with model uncertainty:
\begin{align}
\label{eq:isolated_deg}
 \hat{X}^{(j)}=X^{(j)}+Z^{(j)},\  j\in\{1,\ldots,M\},
\end{align}
where $Z^{(j)}\sim {\cal N}_{\mathbb{C}}(0,\sigma^{(j)2})$ is 
additive Gaussian noise of known variance $\sigma^{(j)2}$.
Note that the noise variance defined here is the estimation error with respect to the ground-truth of the clean speech.
It depends on the uncertainty of the preceding model, and thus calculating this variance $\sigma^{(j)2}$ directly can be challenging.
Hence, we propose an approximation for $\sigma^{(j)2}$ based on a metric related to the target-to-interference ratio (TIR) in the SS task:
\begin{align}
 &\sigma^{(j)} = \min\left(S(\Phi,\hat{X}^{(j)}),\ \sigma^{(j)}_{\mathrm{min}}\right),
 \,j\in\{1,\ldots,M\},\label{eq:sigmoid_var}\\
 &\mathrm{where}\ S(\Phi,\hat{X}^{(j)})=\dfrac{\alpha}{1+\exp (-\beta |\Phi - \hat{X}^{(j)}|)} - \gamma. \nonumber
\end{align}
Here, $\alpha$, $\beta$, and $\gamma$ are hyperparameters to control how the TIR affects the variance, and $\sigma^{(j)}_{\mathrm{min}}$ is a positive parameter to prevent $\sigma^{(j)}$ from becoming negative.
Note that $\sigma^{(j)}$ is calculated for each TF bin $(k, l)$ on the spectrograms.
By approximating the desired variance $\sigma^{(j)2}$ for each TF bin by utilizing the function of the estimated interference (as in Eq.~(\ref{eq:sigmoid_var}), which is related to TIR), Diffiner for SS becomes feasible.

In addition to the design of $\sigma^{(j)}$, we need to consider the design of Eq.~(\ref{eq:isolated_deg}).
Since this equation is built for each $j$th speaker independently, it ignores the outputs of the preceding models, i.e., the relationship among all speakers. 
Inspired by Bayesian Annealed SIgnal Source (BASIS) separation~\cite{jayaram2020source}, we propose using the mixture signal for conditioning the generation process to address this concern. 

BASIS is a diffusion-based separation method for images that takes a mixture image as an observation and uses Langevin dynamics to separate it into individual images. Motivated by the Bayesian Annealing approach in BASIS, we introduce a mechanism in our sampling process where the weight of the observation is gradually increased. This is particularly important for SS, as placing too much emphasis on the observation (mixture signal or outputs from existing SS methods) at the initial stage may amplify errors. When we gradually increase the influence of the observation, we can maintain the naturalness of the generated samples in the early stages by relying more on the prior distribution of the diffusion model, and then strengthen the observational constraints in the later stages to better align each speaker’s audio track with the mixture signal.\looseness=-1

For SS, we combine a mixture signal $\bm{\Phi}$ with the estimates of the preceding models $\hat{X}^{(j)}$ to obtain a new observation. 
This enables us to fully leverage the available information and generate a sample that is faithful to the reference signal. 
For example, for the two-speaker case ($M=2$), the observation in Eq.~(\ref{eq:isolated_deg}) becomes
\begin{align}
\label{eq:two_obs}
&\left\{ \,
 \begin{aligned}
 &\Phi = X^{(1)} + X^{(2)} + Z^{(\varphi)} \\
 &\hat{X}^{(1)} = X^{(1)} + Z^{(1)} \\
 &\hat{X}^{(2)} = X^{(2)} + Z^{(2)}
 \end{aligned}
 \right.,
\end{align}
where $Z^{(\varphi)}$ is additive Gaussian noise of known variance $\sigma^{(\varphi)2}$ for the mixture observation $\Phi$. 
Throughout the subsequent experiments, $\sigma^{(\varphi)2}$ is fixed at 1.

Let $\bm{\Psi}\in\mathbb{C}^{(M+1)\times d_k\times d_l}$ be the observation matrix where the $(k,l)$-component is represented as 
\begin{align}
    \Psi = \begin{bmatrix}
    \Phi & \hat{X}^{(1)} & \cdots & \hat{X}^{(M)}
    \end{bmatrix}^\top.
    \label{eq:shared_deg}
\end{align}
We call Diffiner for SS based on Eq.~{(\ref{eq:isolated_deg})} the \emph{isolated model}, and the one based on Eq.~(\ref{eq:shared_deg}) the \emph{shared model}.

The proposed update is summarized in Algorithm~\ref{alg:proposal_alg} (see ``case \texttt{ss}''), 
and the outline of Diffiner for SS using its update is summarized in Fig.~\ref{fig:overview}.
We first apply singular value decomposition (SVD) to the observation matrix $\bm{\Psi}$ consisting of the input mixture $\bm{\Phi}$ and $M$ speakers' STFT coefficients $\hat{X}^{(j)}$ $(j=1, 2, \cdots, M)$ that are separated by the preceding model.
Then, running the learned Diffiner's reverse process by using the obtained $\bar{\bm{\Psi}}$ under the proposed rule of the shared DDRM update (see the yellow part in Fig.~\ref{fig:overview}), each of the individual speech sources is refined.

\section{Experiments}
\label{sec:exp}
As discussed in the previous sections, we modified Diffiner so that it can be applied to both SE and SS by training the DDPM just once on only clean speech.
Specifically, we trained the model by using the spectrograms of only clean speech from Wall Street Journal (WSJ0)\footnote{\url{https://doi.org/10.35111/ewkm-cg47}}, which consists of approximately 30 hours of speech recordings in total.
This is because Diffiner does not require any noisy signals or noise-only signals in training, as discussed in Sec.~\ref{sec:proposal}.
In other words, we can equalize the dataset to train both Diffiners for SE and SS.
The Hann window size, hop size, and number of time frames were set to $512$, $256$, and $256$, respectively. 
By truncating the direct current (DC) component, we treated the spectrograms as $256 \times 256$ tensors with two channels corresponding to the real and imaginary parts of a complex value.
For all Diffiners, we employed a U-Net-based architecture similar to that used by Dhariwal et al.~\cite{dhariwal2021diffusion}.
Since the size of our target spectrogram is $256 \times 256$, we implemented a $256 \times 256$ U-Net model from the authors' official repository, i.e., OpenAI guided-diffusion\footnote{ \url{https://github.com/openai/guided-diffusion} }.
Please note that we changed the numbers of both input and output channels from three to two, i.e., from RGB to real and imaginary valued STFT coefficients.
An example of Diffiner implementation is available at our repository\footnote{\url{https://github.com/sony/diffiner}}.

We trained the model on a single NVIDIA A100 GPU ($40$ GB memory) for $3.5 \times 10^5$ steps, which took about three days. 
We used the Adam optimizer~\cite{kingma2014adam} with a learning rate of $\num{1.0e-3}$, and the batch size of eight. 
Following a previous study~\cite{song2020improved}, we took an exponential moving average of the model weights with a decay of $0.9999$ to be used for inference. 

For inference, we first conducted a grid search for the parameters required in Diffiner: $\eta_a$ and $\eta_b$ in Algorithm~\ref{alg:proposal_alg}.
Specifically, we searched the parameter space in $[0, 1]$ with a step size of $0.1$ for both $\eta_a$ and $\eta_b$. 
Furthermore, $\lambda$, $\delta$, and $R$ were set to $\lambda=1.0$, $\delta=\num{1.0e-5}$, and $R=\sigma^2_{T-1}\approx97$. 
In addition, $\alpha, \beta$, and $\gamma$ in Eq.~(\ref{eq:sigmoid_var}) were set to $2.0, 2.0$ and $0.8$, respectively. 
These values were determined through preliminary experiments.
Then, we ran all instances of Diffiner with $T=200$.
Note that a more detailed parameter search is described in Appendix~\ref{app:hyper_params}.\looseness=-1

After that, to confirm the versatility and validity of our method, we applied the obtained Diffiners to diverse results from SE and SS.
Specifically, we used a number of SE and SS methods, each having different characteristics, to output some results and then refined those results by applying Diffiner.

We used five metrics to quantitatively evaluate the performance\footnote{To compute these metrics, we utilized the latest versions of the following public codes as of March 2025:\\
$\cdot$\url{https://github.com/sigsep/sigsep-mus-eval}\\
$\cdot$\url{https://github.com/gabrielmittag/NISQA}\\
$\cdot$\url{https://github.com/microsoft/DNS-Challenge}}: SI-SDR~\cite{le2019sdr}, WB-PESQ~\cite{rix2001perceptual}, ESTOI~\cite{taal2010short}, non-intrusive speech quality assessment (NISQA)~\cite{mittag2021nisqa}, and the overall (OVRL) metric of the deep noise challenge mean opinion score (DNSMOS) P.835~\cite{reddy2022dnsmos}.
The first three of these metrics, which require the pair of a target signal and the corresponding reference source, have recently been used for evaluating the performance of speech processings~\cite{richter2023speech, scheibler2022diffusion}.
We therefore focus on improving the final two metrics, i.e., NISQA and OVRL, in our experiments.
This is because some recent methods, including Diffiner, have generative processing parts whose outputs may contain some generated parts, and thus they do not always match the corresponding references.
Hence, the reference-based metrics calculated from such results may produce unreasonably poor scores even if the results sound better in perceptual quality than the input.
In contrast, NISQA and OVRL can predict the mean opinion score (MOS) of a target signal without a corresponding reference source.
Thus, we predicted that they would be especially suitable for evaluating generative model-based results.

However, the straightforward way to evaluate perceptual quality is to conduct an actual human listening test.
Therefore, we additionally utilized the MUltiple Stimuli with Hidden Reference and Anchor (MUSHRA) methodology~\cite{bureau2001method} to conduct a human listening test comparing signal examples before and after applying Diffiner.
We used the web-based MUSHRA platform available at the public GitHub repository\footnote{\url{https://github.com/audiolabs/webMUSHRA}}.
In our MUSHRA-based human listening test, five stimuli comprising four processed signals and one hidden reference identical to the original signal, were presented for each test item.
Note that no anchor signal was included since the MUSHRA-based standard anchor (such as a signal filtered by a low-pass filter) may not work well.
As discussed in \cite{lemaguer25_ssw}, depending on the experimental target, such a simple anchor is substantially different from other stimuli and easy to distinguish.
In our case, since almost all stimuli were relatively high-quality processed signals, we excluded such simple MUSHRA-based anchors applied by a low-pass filter.
Twelve listeners participated in the experiment, all of whom were engineers involved in audio research and experienced in critical listening tests.
Ten audio samples were randomly selected from the test dataset and evaluated by all listeners.\looseness=-1

\subsection{Speech enhancement}
\label{subsec:speech_enhancement}
As discussed in Sec.~\ref{sec:intro}, we previously proposed Diffiner for only SE~\cite{sawata23_interspeech} and evaluated it on the famous small dataset, VoiceBank-DEMAND (VBD)~\cite{voicebankdemand}.
To confirm that Diffiner can work with the large-scale data regime, we built a new version on a larger dataset and examined its effectiveness through not only quantitative but also qualitative evaluations.
Specifically, as described at the beginning of Sec.~\ref{sec:exp}, we newly trained Diffiner on WSJ0 which consists of approximately thirty hours of clean speech---significantly more than the total duration of clean speech in the previous Diffiner on VBD, which was approximately nine hours.

\subsubsection{Preceding methods}
\label{subsec:preceding_speech_enhancement}
To assess the versatility of Diffiner in the SE task, we paired it with three different SE methods: i) a classical Wiener filter~\cite{wiener_org}, ii) a non-generative state-of-the-art (SOTA) method, MP-SENet~\cite{lu2023mp}, and iii) a generative SOTA method, StoRM~\cite{lemercier2022storm}.
The learning-based methods, i.e., MP-SENet and StoRM, were trained on the same dataset, WSJ0+CHiME, which was simulated by using clean speech from WSJ0 and noise from the dataset of the 3rd Computational Hearing in Multisource Environments challenge (CHiME-3) \cite{barker2015third}.
Each speech and noise sample was randomly selected from these datasets and mixed so that the mixed noisy signals followed a uniform distribution of SNRs [--6dB, 14dB].
Note that the authors of StoRM have published the script used to create WSJ0+CHiME on their official GitHub repositories\footnote{ MP-SENet: \url{https://github.com/yxlu-0102/MP-SENet} and StoRM: \url{https://github.com/sp-uhh/storm}}.

\subsubsection{Results}
\label{subsec:results_se}

\begin{table}[t]
\setlength\tabcolsep{0pt}
\caption{
    Results of Diffiner on speech enhancement and speech separation tasks.
    Note that the reference-based metrics require ground-truth speech as references, whereas the reference-free metrics do not require ground-truth speech.
    For all metrics, higher is better.
}
 \label{table:results_methods}
 \begin{subtable}[t]{\columnwidth}
\subcaption{Speech enhancement on WSJ0+CHiME dataset.}
 \label{table:speech_enhancement_wsj}
 \centering
 \small
 \begin{tabular*}{\columnwidth}{@{\extracolsep{\fill}}lccccc}
 \toprule
  & \multicolumn{3}{c}{Reference-based} & \multicolumn{2}{c}{Reference-free} \\ \cmidrule{2-4} \cmidrule{5-6}
  & SI-SDR$\uparrow$  & PESQ$\uparrow$ & ESTOI$\uparrow$ & NISQA$\uparrow$ & DNSMOS$\uparrow$ \\
 \midrule
 Clean & -- & -- & -- & 4.20 & 3.36\\
 Mixture & 5.19 & 1.36 & 0.64 & 2.12 & 2.30\\
 \midrule
 Wiener filter~~\cite{wiener_org} & 8.55 & 1.59 & 0.67 & 2.45 & 2.44\\
 \multicolumn{1}{l}{\ \ \ w/ Diffiner} & 12.4 & 1.79 & 0.79 & 4.21 & 3.00 \\
 \multicolumn{1}{l}{\ \ \ w/ Diffiner+} & \textbf{12.4} & \textbf{1.86} & \textbf{0.80} & \textbf{4.22} & \textbf{3.03}\\
 \midrule
 MP-SENet~\cite{lu2023mp} & \textbf{17.5} & \textbf{3.17} & \textbf{0.92} & 4.02 & 3.19\\
 \multicolumn{1}{l}{\ \ \ w/ Diffiner} & 16.8 & 2.65 & 0.91 & 4.52 & 3.22 \\
 \multicolumn{1}{l}{\ \ \ w/ Diffiner+} & 16.6 & 2.68 & 0.91 & \textbf{4.53} & \textbf{3.23}\\
 \midrule
 StoRM~\cite{lemercier2022storm} & 15.5 & \textbf{2.56} & 0.88 & 4.35 & 3.18\\
 \multicolumn{1}{l}{\ \ \ w/ Diffiner} & \textbf{16.3} & 2.48 & 0.89 & 4.55 & 3.20 \\
 \multicolumn{1}{l}{\ \ \ w/ Diffiner+} & 16.2 & 2.51 & \textbf{0.89} & \textbf{4.57} & \textbf{3.21}\\
 \bottomrule
 \end{tabular*}
\end{subtable}

\vspace{1em}

\begin{subtable}[t]{\columnwidth}
\caption{ 
    Speech separation results on WSJ0-2mix dataset.
    The underlined scores mean the method with Diffiner was not the best overall, but that it outperformed the corresponding original SS method without Diffiner.
}
 \label{table:results_sep}
 \centering
 \small
 \begin{tabular*}{\columnwidth}{@{\extracolsep{\fill}}lccccc}
 \toprule
  & \multicolumn{3}{c}{Reference-based} & \multicolumn{2}{c}{Reference-free} \\ \cmidrule{2-4} \cmidrule{5-6}
  & SI-SDR$\uparrow$ & PESQ$\uparrow$ & ESTOI$\uparrow$ & NISQA$\uparrow$ & DNSMOS$\uparrow$ \\
 \midrule
 Clean & -- & -- & -- & 3.53 & 3.29\\
 Mixture & 2.52 & 1.22 & 0.67 & 2.46& 2.78\\
 \midrule
 Deep Clustering~\cite{hershey2016deep} & 8.90 & 1.63 & \textbf{0.67} & 1.91 & 2.74\\
 \multicolumn{1}{l}{\ \ \ w/ Diffiner} & 8.88 & 1.31 & 0.65 & \textbf{3.41} & \textbf{3.03}\\
 \multicolumn{1}{l}{\ \ \ Blended} & \textbf{9.00} & \textbf{1.68} & \textbf{0.67} & \underline{2.54} & \underline{3.01} \\
 \midrule
 Conv-TasNet~\cite{luo2019conv} & \textbf{18.5} & 3.06 & 0.80 & 3.08 & 3.27\\
 \multicolumn{1}{l}{\ \ \ w/ Diffiner} & 17.6 & 1.74 & 0.80 & \textbf{3.99} & \textbf{3.31}\\
 \multicolumn{1}{l}{\ \ \ Blended} & \textbf{18.5} & \textbf{3.07} & \textbf{0.81} & \underline{3.35} & \textbf{3.31}\\
 \midrule
 DPRNN~\cite{luo2020dual} & \textbf{20.0} & 3.46 & 0.84 & 3.48 & 3.27 \\
 \multicolumn{1}{l}{\ \ \ w/ Diffiner} & 18.7 & 1.85 & 0.82 & \textbf{4.24} & \textbf{3.33} \\
 \multicolumn{1}{l}{\ \ \ Blended} & \textbf{20.0} & \textbf{3.49} & \textbf{0.85} & \underline{3.59} & \underline{3.32}\\
 \midrule
 Sepformer~\cite{subakan2021attention} & 23.0 & 3.86 & \textbf{0.97} & 3.53 & 3.29\\
 \multicolumn{1}{l}{\ \ \ w/ Diffiner} & 22.0 & 2.94 & 0.96 & \textbf{3.76} & \textbf{3.34}\\
 \multicolumn{1}{l}{\ \ \ Blended} & \textbf{23.1} & \textbf{3.87} & \textbf{0.97} & \underline{3.63} & \underline{3.31} \\
 \bottomrule
 \end{tabular*}
\end{subtable}
\end{table}

\begin{figure}[t]
 \centering
 \begin{subfigure}{0.24\textwidth}
    \includegraphics[width=\columnwidth]{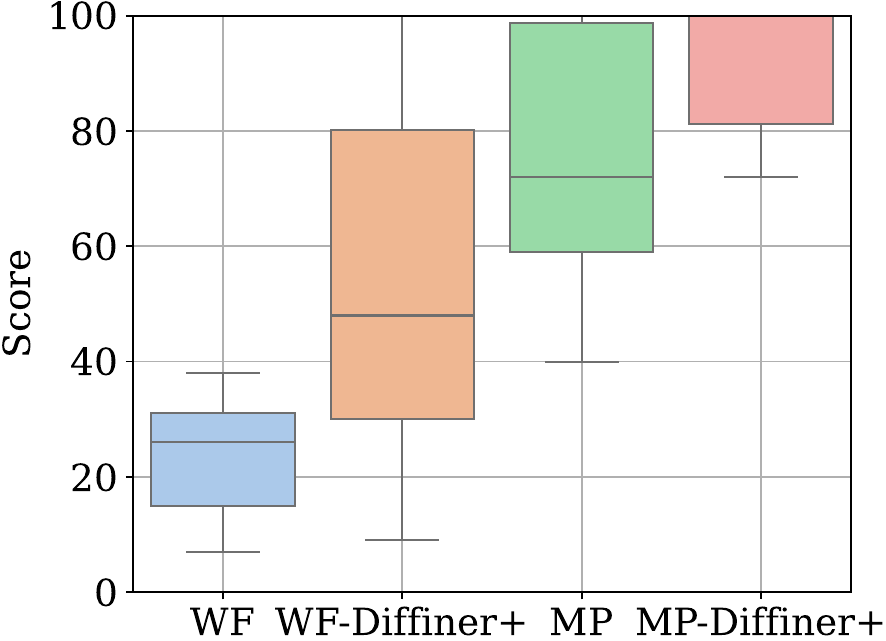}
    \caption{Results for speech enhancement. WF: Wiener Filter, MP: MP-SENet.}
 \end{subfigure}
 \hfill
 \begin{subfigure}{0.24\textwidth}
    \includegraphics[width=\columnwidth]{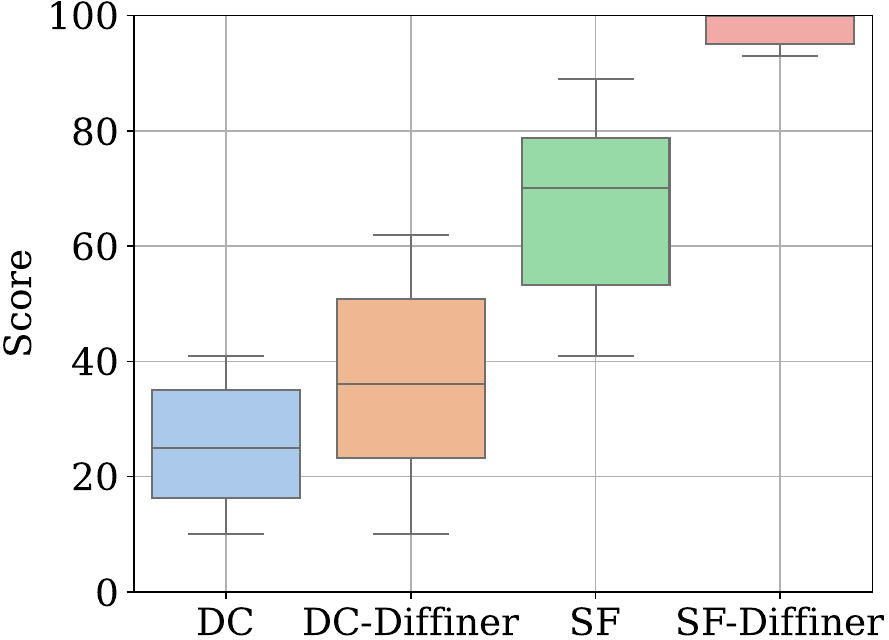}
    \caption{Results for speech separation. DC: Deep Clustering, SF: Sepformer.}
 \end{subfigure}  
 \caption{
    Boxplots comparing the results of MUSHRA subjective listening test for speech enhancement (left) and speech separation (right).
    Twelve participants took part in the experiment.
 }
\label{fig:mushra}
\end{figure}

The quantitative results are summarized in Table~\ref{table:speech_enhancement_wsj}. 
As we can see, NISQA and DNSMOS improved for all methods when Diffiner or Diffiner+ was used. 
This indicates that Diffiner succeeded in improving the perceptual quality of the audio no matter which module preceded it.
In particular, after applying it, the NISQA and DNSMOS scores were superior even to those of the clean signal.
This implies that our refiner might have succeeded in generating some plausible parts of the spectrogram not included in the original recorded clean signal.
This may have been due to an imperfect recording environment featuring, for example, background noise, and thus it would have had undesirable noise and artifacts that degraded the scores.
In contrast, Diffiner and Diffiner+ marginally degraded the reference-based metrics. 
This is because, as discussed previously, the refined results would have contained generated parts that may have been plausible but did not match the reference. 
Note that more detailed analysis and discussion in terms of noise type and the difference between DNSMOS and NISQA are presented in Appendices~\ref{app:dns_nisqa} and \ref{app:noise_type_gender}.\looseness=-1

In our MUSHRA-based human listening test for SE,
Wiener Filter (WF) and MP-SENet (MP) were used as the preceding models, and all participants listened to and compared the results of before and after applying Diffiner.
Figure~\ref{fig:mushra}(a) indicates that the human listening test confirmed that Diffiner could improve the perceptual quality by replacing the degraded parts of the spectrogram with the newly generated parts.
Namely, this would have been reflected in improvements to not only reference-free metrics but also human listening. 

\subsection{Speech separation}
\label{subsec:speech_separation}

\subsubsection{Preceding methods}
\label{subsec:preceding_speech_separation}
We experimented with various types of preceding SS methods:
a traditional method called Deep Clustering~\cite{hershey2016deep}\footnote{
\url{https://github.com/JusperLee/Deep-Clustering-for-Speech-Separation}
},
a CNN-based method called Conv-TasNet~\cite{luo2019conv}\footnote{ \url{https://zenodo.org/record/3862942#.ZDSjY3bP02w} }, 
an RNN-based method called Dual-Path RNN (DPRNN)~\cite{luo2020dual}\footnote{ \url{https://zenodo.org/record/3903795#.ZDSjfXbP02w} }, 
and a transformer-based SOTA method called Sepformer~\cite{subakan2021attention}\footnote{ \url{https://huggingface.co/speechbrain/sepformer-wsj02mix} }.
All of these were trained on the same dataset, WSJ0-2mix~\cite{hershey2016deep}.
Note that all signals in WSJ0-2mix were resampled to $8$ kHz.\looseness=-1





\subsubsection{Preliminary}
\label{subsec:pre_results}
As discussed in Sec.~\ref{subsec:refiner_sep}, Diffiner for SS has two types of model design equation: one for \emph{isolated} models and one for \emph{shared} models (see Eqs.~(\ref{eq:isolated_deg}) and (\ref{eq:shared_deg})).
Furthermore, we proposed an approximation of the variance $\sigma^{(j)}$ to run Diffiner's forward and backward processes for SS, called \emph{sigmoid}-based design (see Eq.~(\ref{eq:sigmoid_var})).
To confirm its validity, we compared it with a simple and naive design, just using a \emph{fixed} degree of power of Gaussian noise.
Hence, as a preliminary experiment, we compared four ways using Diffiner for SS, i.e., as an \emph{isolated} or as a \emph{shared} model with a \emph{sigmoid-}based design of the observation noise or naive design simply using a \emph{fixed} power level of Gaussian noise, and chose the best for the following comprehensive experiments.
In summary, the four ways compared in this preliminary experiment are: i) \emph{isolated} model with \emph{fixed} noise, ii) \emph{isolated} model with \emph{sigmoid}-based noise, iii) \emph{shared} model with \emph{fixed} noise, and iv) \emph{shared} model with \emph{sigmoid}-based noise.
Note that Conv-TasNet was used as the preceding SS method.
%
\begin{table}[t]
\setlength\tabcolsep{0pt}
\caption{
    Results of preliminary experiments on Conv-TasNet refined by using Diffiner in four ways, i.e., isolated/shared model with fixed-/sigmoid-based noise (see Eqs.~(\ref{eq:isolated_deg})--(\ref{eq:shared_deg}) in Sec.~\ref{subsec:refiner_sep}).
}
 \label{table:pre_results}
 \centering
 \small
 \resizebox{1.0\columnwidth}{!}{%
     \begin{tabular*}{\columnwidth}{@{\extracolsep{\fill}}lccccc}
     \toprule
      & SI-SDR$\uparrow$ & PESQ$\uparrow$ & ESTOI$\uparrow$ & NISQA$\uparrow$ & DNSMOS$\uparrow$ \\
     \midrule
     Isolated\_fixed & 16.4 & 1.37 & 0.77 & 3.92 & 3.28 \\
     Isolated\_sigmoid & 17.3 & 1.52 & \textbf{0.80} & 3.76 & \textbf{3.32} \\
     Shared\_fixed & \textbf{17.6} & 1.66 & \textbf{0.80} & 3.91 & 3.30 \\
     Shared\_sigmoid & \textbf{17.6} & \textbf{1.74} & \textbf{0.80} & \textbf{3.99} & 3.31 \\
     \bottomrule
     \end{tabular*}%
 }
\end{table}

As shown in Table~\ref{table:pre_results}, the shared model with the sigmoid function-based design of the noise performed the best.
Therefore, all of the following experimental results on Diffiner for SS used this configuration.\looseness=-1

\begin{figure}[t]
 \centering
 \includegraphics[scale=0.44]{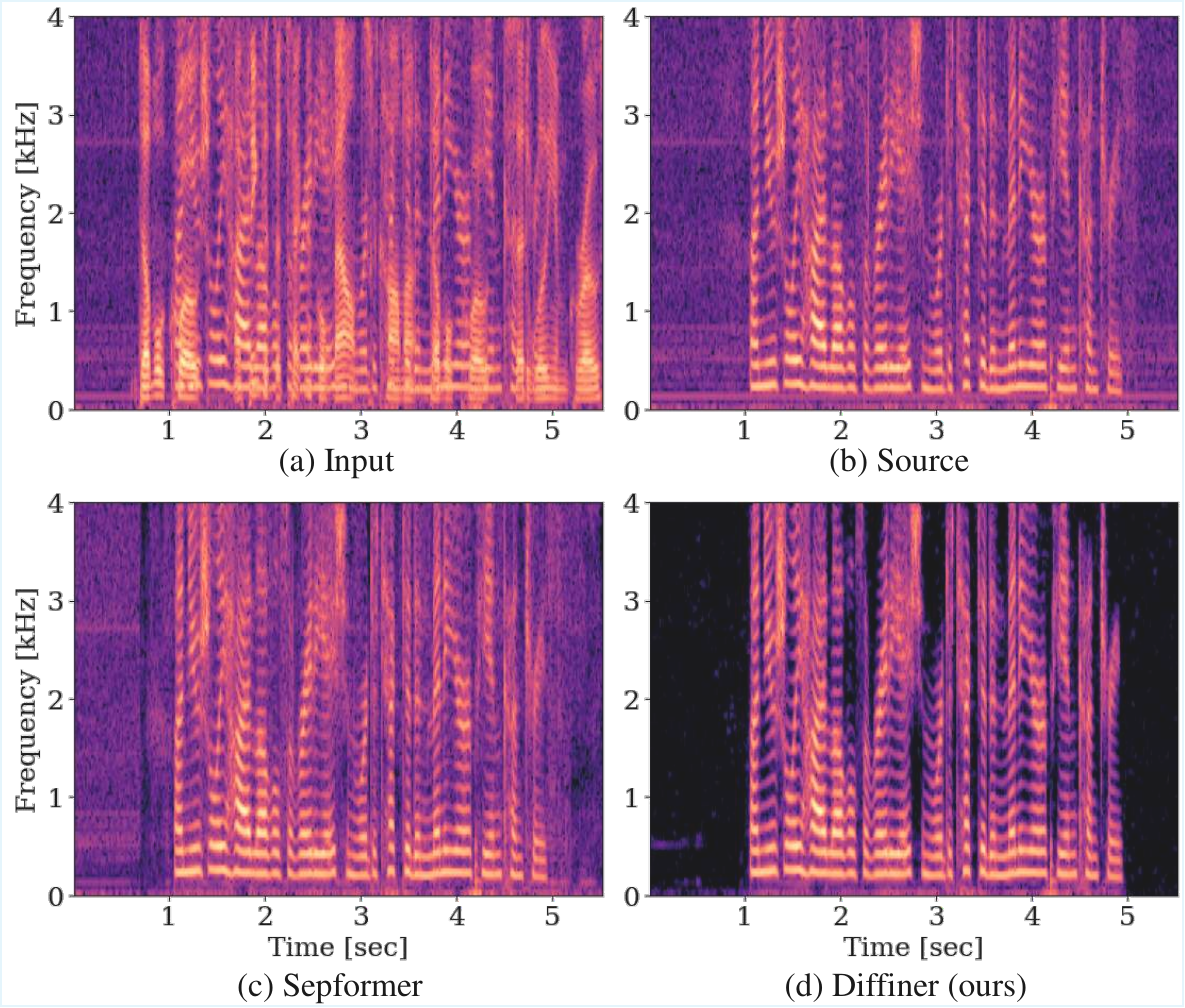}
 \caption{
    Comparison of spectrograms.
 }
 \label{fig:processed_example}
\end{figure}

\subsubsection{Results}
\label{subsec:separation_results}
The quantitative results are summarized in Table~\ref{table:results_sep}.
As shown, applying Diffiner improved the perceptual metrics, i.e., NISQA and DNSMOS, of all of the preceding SS methods.
Moreover, although the reference-based metrics, i.e., SI-SDR, PESQ, and ESTOI, were marginally degraded, the results were nonetheless convincing because Diffiner does not aim to restore the corresponding reference signals but rather to refine the pre-processed results resulting into natural-looking spectrograms.
Thus, the scores of reference-free metrics rather than reference-based ones were indeed improved.
Note that more detailed analysis and discussion in terms of speaker gender and the difference between DNSMOS and NISQA are presented in Appendices~\ref{app:dns_nisqa} and \ref{app:noise_type_gender}.

\begin{figure}[t]
 \centering
 \includegraphics[width=\columnwidth]{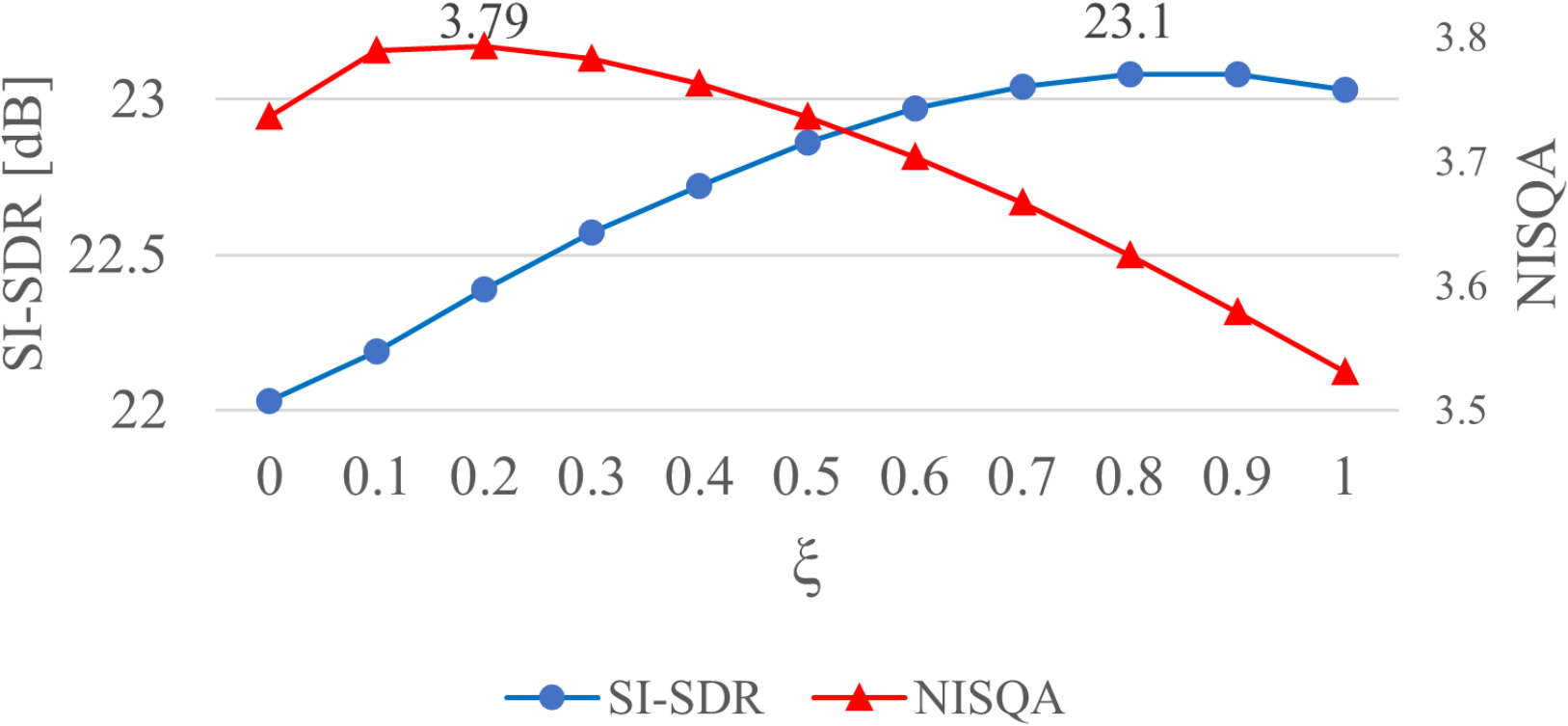}
 \caption{
    Change in reference-based and reference-free metrics, i.e., SI-SDR and NISQA, by changing the blending weight $\xi$.
    The blended signal $\tilde{\bm{x}}$ was calculated by simply adding the weighted preceding method's output $\bm{x}_0$ and Diffiner's output $\hat{\bm{x}}$, 
    i.e., $\tilde{\bm{x}} = \xi \bm{x}_0 + (1-\xi) \hat{\bm{x}}$.
}
 \label{fig:blend_results}
\end{figure}

Furthermore, we found a blending strategy.
Specifically, we could improve both the reference-based and reference-free metrics by simply blending weighted results before and after refinement.
In fact, all of the blended results shown in Table~\ref{table:results_sep} are bolded or underscored (see the lines labeled ``Blended''), meaning that, although they might not be the best scores, they were always better than those of the corresponding preceding method alone.
Note that we conducted a grid search for the weighting parameter as shown in Fig.~\ref{fig:blend_results}, and the best trade-off between the reference-based and reference-free metrics was adopted for each preceding method.

\begin{figure*}[t]
  \centering
  \begin{subfigure}{0.24\textwidth}
    \centering
    \includegraphics[width=\linewidth]{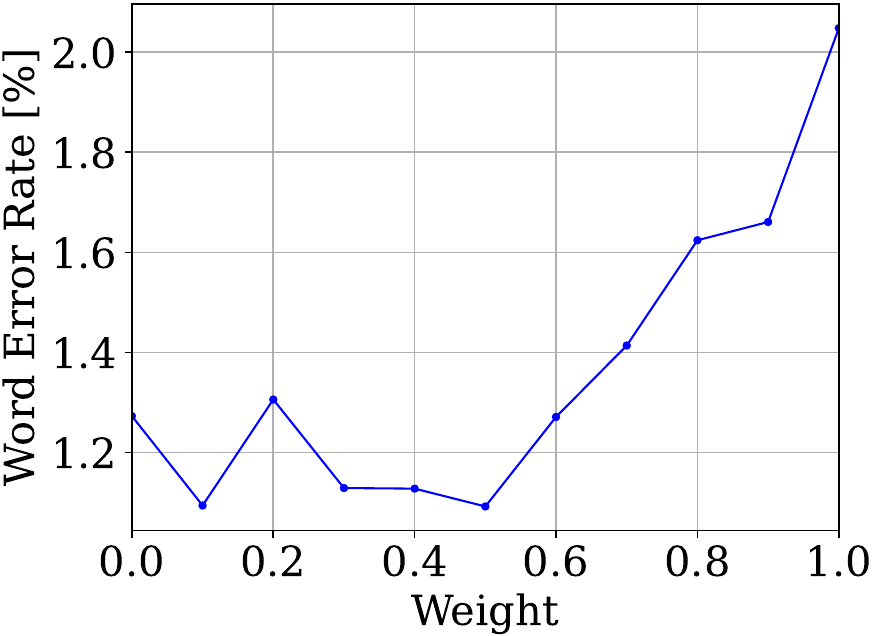}
    \caption{OpenAI Whisper~\cite{radford2023robust}}
    \label{fig:openai}
  \end{subfigure}
  \hfill 
  \begin{subfigure}{0.24\textwidth}
    \centering
    \includegraphics[width=\linewidth]{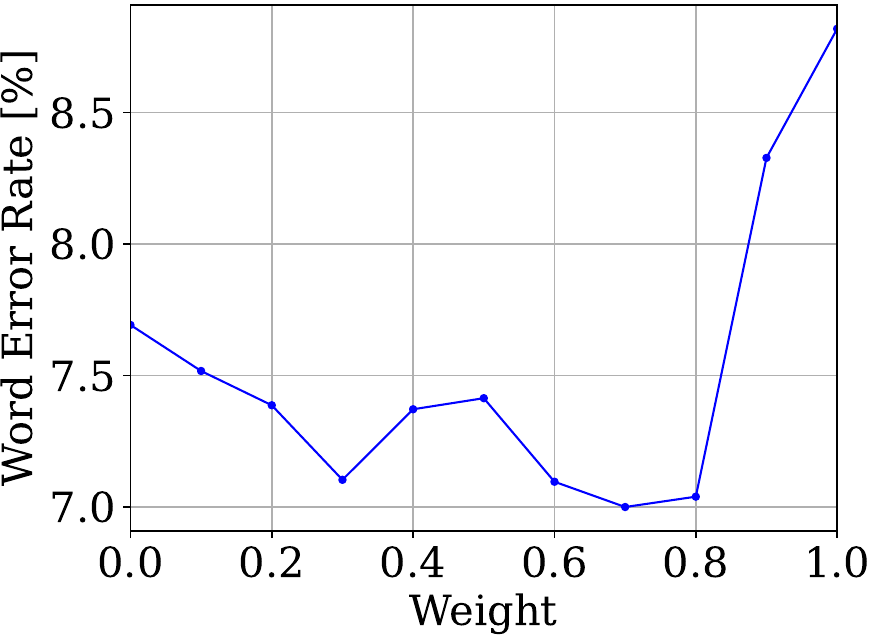}
    \caption{Fast Whisper~\cite{machavcek2023turning}}
    \label{fig:whisper}
  \end{subfigure}
  \hfill
  \begin{subfigure}{0.24\textwidth}
    \centering
    \includegraphics[width=\linewidth]{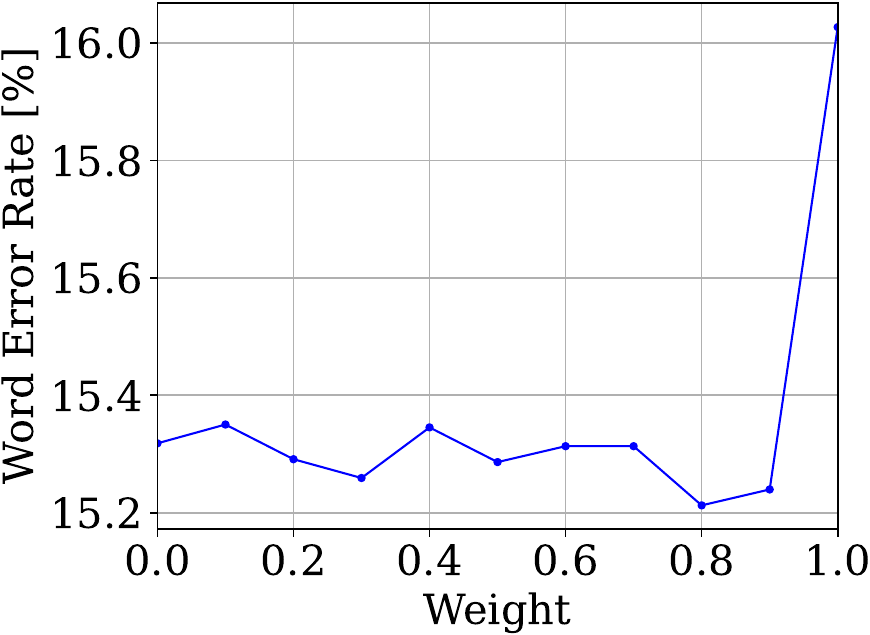}
    \caption{Wav2Vec~\cite{baevski2020wav2vec}}
    \label{fig:wav2vec}
  \end{subfigure}
  \hfill
  \begin{subfigure}{0.24\textwidth}
    \centering
    \includegraphics[width=\linewidth]{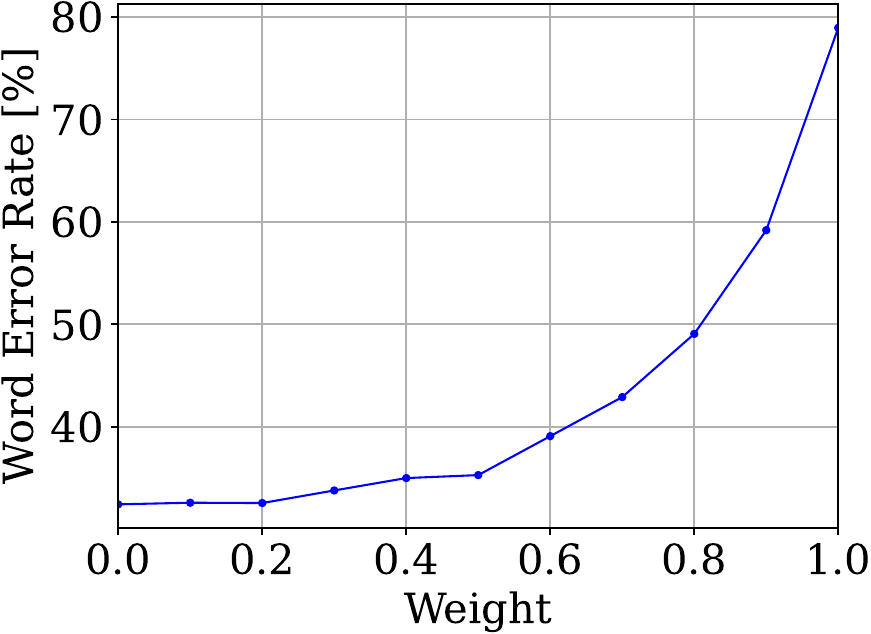}
    \caption{Sphinx~\cite{lee1990overview}}
    \label{fig:sphinx}
  \end{subfigure}

  \caption{
    Word Error Rate evaluation using different ASR models for blended outputs.
    The horizontal axis corresponds to the blending coefficient $\xi$, ranging from 0.0 to 1.0.
    The blended signal $\tilde{\bm{x}}$ was obtained as $\tilde{\bm{x}} = \xi \bm{x}_0 + (1-\xi) \hat{\bm{x}}$, where $\bm{x}_0$ and $\hat{\bm{x}}$ are respectively the outputs of Dual-path RNN (DPRNN) and Diffiner.
  }
  \label{fig:wer_blend}
\end{figure*}

In addition to improving the scores of SI-SDR and NISQA, 
we also conducted experiments to verify the blending approach on automatic speech recognition (ASR) which is known as one of the possible downstream tasks following speech enhancement and separation.
First, we confirmed the influence of Diffiner without the blending approach in terms of phoneme similarity.
As shown in Appendix~\ref{app:phoneme_sim}, Diffiner neither improved nor degraded phoneme similarity compared to the original values (see Fig.~\ref{fig:phoneme_similarity_rev3}).
Therefore, to make use of Diffiner for the ASR task, simply using the output of Diffiner may not be the best way.
Next, we applied several acoustic models (AMs) of ASR, \emph{OpenAI Whisper}~\cite{radford2023robust}, \emph{Fast Whisper}~\cite{machavcek2023turning}, \emph{Wav2Vec}~\cite{baevski2020wav2vec}, and \emph{Sphinx}~\cite{lee1990overview}\footnote{All AMs were cited from a public GitHub repository (\url{https://github.com/Uberi/speech_recognition}.} to the speech signals refined by Diffiner with the blending approach.
We then changed the blending coefficient $\xi$ from 0.0 to 1.0 in increments of 0.1 and calculated word error rate (WER).
The results are shown in Fig.~\ref{fig:wer_blend}. 
In contrast to the relationship between the blending coefficient $\xi$ and the scores of speech quality, i.e., SI-SDR and NISQA (see Fig.~\ref{fig:blend_results}), the line of WER versus $\xi$ was not a single-peaked curve.
Especially, the best blending coefficients $\xi$ for WERs were different depending on which AM was used, even though their task was common, i.e., ASR.
Therefore, the optimal blending might be diverse depending on what the final purpose is (e.g., best SNR, best perceptual quality, best performance on the downstream task); furthermore, it depends on the detailed parts like the above AM rather than the entire system.
One of the most straightforward ways would be to train the refinement model with the criterion related to the final purpose, e.g., optimizing the model so that WER is the lowest, but this is laborious and not time-efficient.
Hence, our idea of blending is considered useful in that it practically allows for obtaining a relatively good output simply by adjusting weights.


Next, we conducted a MUSHRA-based human listening test in the same manner as the SE experiments.
Deep clustering (DC) and Sepformer (SF) were employed as the preceding models, and 12 participants listened to and compared the results of before and after applying Diffiner.
The results, summarized in Fig.~\ref{fig:mushra}(b), indicate that methods applying Diffiner outperformed the corresponding methods that did not apply it by a significant margin.
In particular, the qualitative results of the refinement were especially improved (see Fig.~\ref{fig:processed_example} as an example).
\\

In summary, it is clear that we can naturally extend Diffiner for SE into a new refiner for SS without any re-training costs because its extension is simply upon inference.
Diffiner improved the perceptual quality, i.e., both reference-free metrics and human listening, whereas the traditional reference-based metrics were slightly degraded.
However, since we found a way to improve both types of metric through the blending strategy, users have the option to use it depending on the downstream task.\looseness=-1

%

\section{Conclusion}
\label{sec:conclusion}
We have proposed a versatile speech refiner based on a diffusion generative model, named Diffiner.
In this study, by deriving a new way of inference for the SS task, we made Diffiner feasible for SS as well as SE.
Since Diffiners for SE and SS are based on the same network trained on only clean speech, users can refine any speech signal pre-processed by any SE and SS model without having to perform additional training.
Therefore, we argue that Diffiner has high versatility and modularity w.r.t. the existing speech processors.
Namely, all we have to do is train a diffusion-based model once by using only clean speech and then refine any speech signals by using Diffiner depending on the desired inference target.
Experimental results showed improvements not only in scores related to human perceptual quality, i.e., reference-free metrics, but also in MUSHRA-based human listening tests.
Furthermore, we found that a simple blending strategy can improve perceptual quality while maintaining the performance of traditional reference-based metrics.

However, the degraded speech signals used in our experiments were still unnatural; speech signals recorded in the wild may contain noise, reverberation, and partially or fully overlapped interfering speeches.
Furthermore, depending on the downstream task, running a system with Diffiner in real time is required (see the analysis regarding computational cost versus performance in Secs.~S-I and S-II of the supplemental material). 
Therefore, in the future, evaluating Diffiner on more realistic scenarios would be a prudent next step.
Then, based on such feasibility testing in the wild, exploiting a unified model that can accept any types of degradation (such as noise, reverberation, and interfering speeches) and subsequently refine them in real time based on generative prior will be preferable and a possible future direction of Diffiner.\looseness=-1



\appendices

\section{Analysis of DNSMOS and NISQA}
\label{app:dns_nisqa}
\begin{figure}[t]
 \centering
  \includegraphics[scale=0.45]{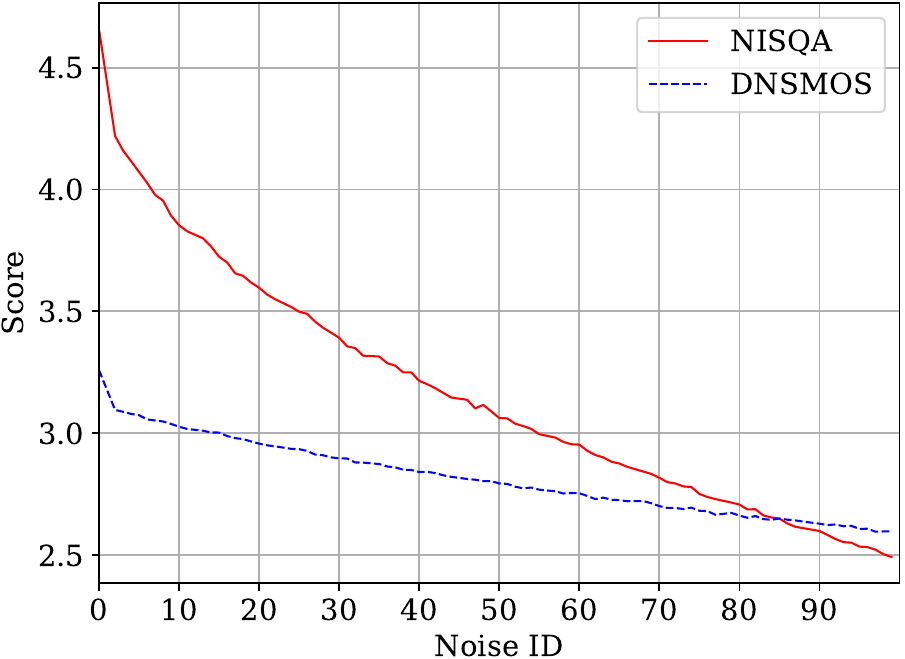}
  \caption{
    NISQA vs. DNSMOS at gradually changing levels of Gaussian noise added to the output of Diffiner.
    The horizontal axis represents the Noise ID, where ``0'' corresponds to the power of background noise of Diffiner's output, and ``100'' corresponds to one of the corresponding original reference signals.
    From Noise \#1 to \#99, Gaussian noise was gradually added so that the RMS of Noise \#100 is equal to the original reference signal's RMS of background noise.
    }
 \label{fig:nisqa_vs_dnsmos}
\end{figure}
As shown in Table~\ref{table:results_methods}, we observed that NISQA can yield higher values compared to the reference, while DNSMOS sometimes scores lower than the reference. 
According to their original papers and backgrounds~\cite{mittag2021nisqa, reddy2022dnsmos}, DNSMOS was proposed to measure how speech processing methods work well on the task of SE.
In the field of SE, achieving the original recording quality, which is often denoted as ``clean speech'' including background noise, is recognized as ideal output.
Therefore, outputting such speech signals including background noise is fine for evaluation on the SE task, and thus DNSMOS would be designed to be insensitive to background noise.
In contrast, the purpose of NISQA is to evaluate the speech quality on voice over internet protocol (VoIP).
In terms of quality evaluation on VoIP, something like Gaussian noise, which is similar to the above background noise, should be penalized because it often harms user experience on VoIP.
Hence, NISQA would be designed to be sensitive to such background noise.

In fact, we observed that Diffiner tends to suppress the leftover background noise from the preceding method because Diffiner was a generative model trained on clean speech only: namely, it can implicitly suppress any sounds except speech.
To confirm the robustness to background noise of NISQA and DNSMOS, we conducted an additional preliminary experiment.
Specifically, we changed the power degree of background noise from Diffiner's output to the corresponding reference speech, i.e., the original recording having background noise, little by little, and we monitored the scores of NISQA and DNSMOS at each point.

The results are shown in Fig.~\ref{fig:nisqa_vs_dnsmos}.
Observing the trends of NISQA and DNSMOS, we can confirm that NISQA changes more sensitively in response to Gaussian noise, which matches our expectations in the aforementioned discussions. 
Since both NISQA and DNSMOS are DNN-based black box models, noise robustness is not the only factor influencing these metrics; however, it is evident that NISQA is more significantly affected by changes in noise. 
In conclusion, we consider that the aforementioned phenomenon was caused by the above difference of metrics' sensitivity to background noise.

\section{Grid Search for Hyper-parameters}
\label{app:hyper_params}
\begin{figure*}[t] 
  \centering
  \begin{subfigure}{0.24\textwidth}
    \includegraphics[width=\linewidth]{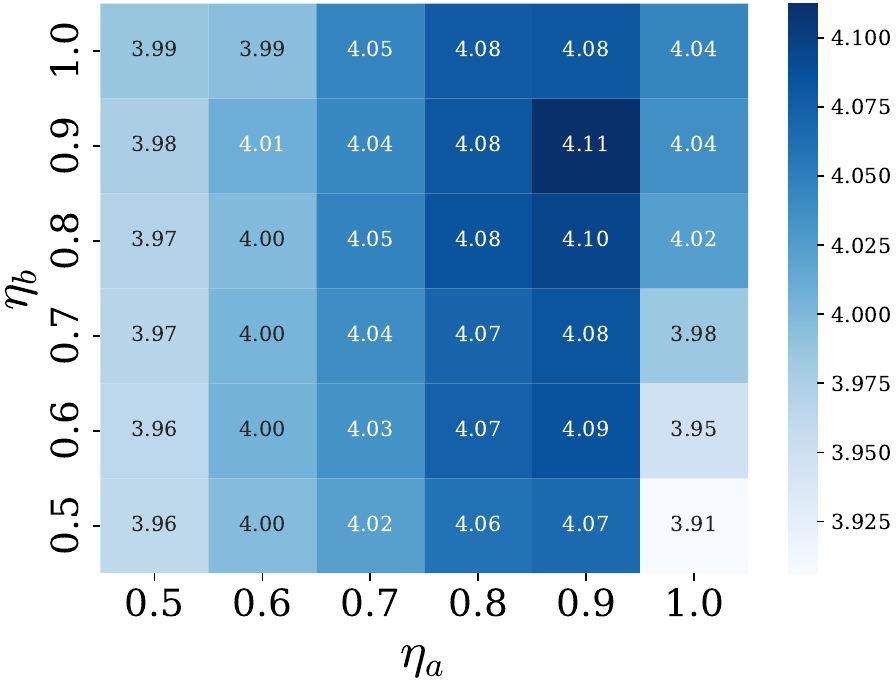}
    \caption{$\eta_a$ and $\eta_b$}
    \label{fig:etas}
  \end{subfigure}
  \hfill 
  \begin{subfigure}{0.24\textwidth}
    \includegraphics[width=\linewidth]{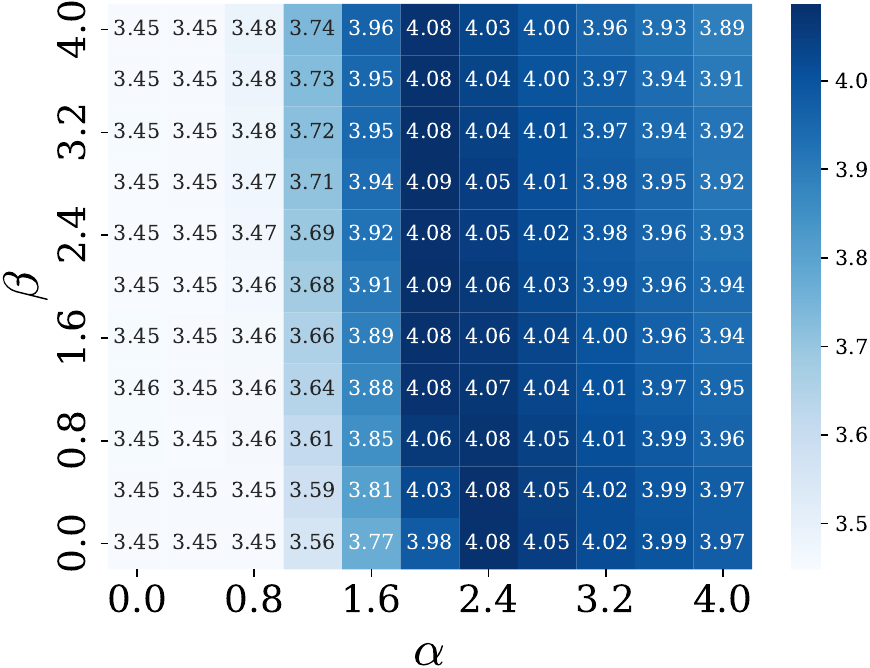}
    \caption{$\alpha$ and $\beta$}
    \label{fig:alpha_beta}
  \end{subfigure}
  \hfill
  \begin{subfigure}{0.24\textwidth}
    \includegraphics[width=\linewidth]{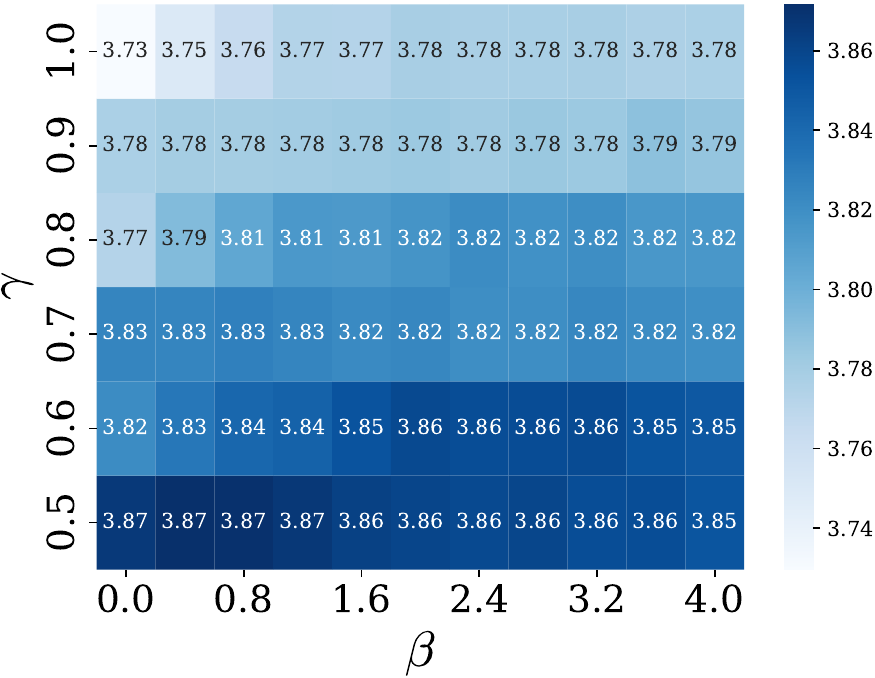}
    \caption{$\beta$ and $\gamma$}
    \label{fig:beta_gamma}
  \end{subfigure}
  \hfill
  \begin{subfigure}{0.24\textwidth}
    \includegraphics[width=\linewidth]{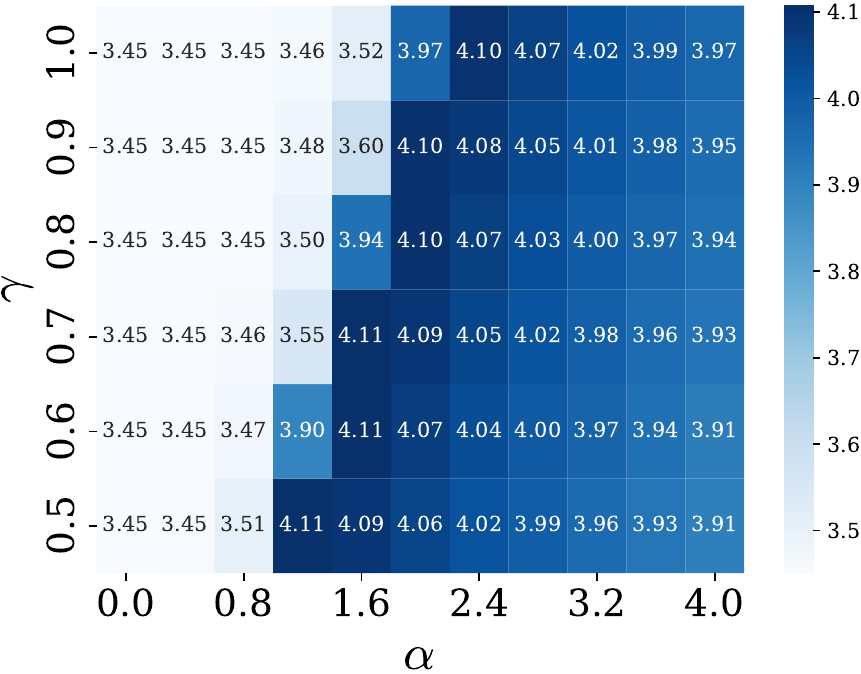}
    \caption{$\alpha$ and $\gamma$}
    \label{fig:alpha_gamma}
  \end{subfigure}

  \vspace{1em} 

  \hspace{+10mm}
  \begin{subfigure}{0.24\textwidth}
    \includegraphics[width=\linewidth]{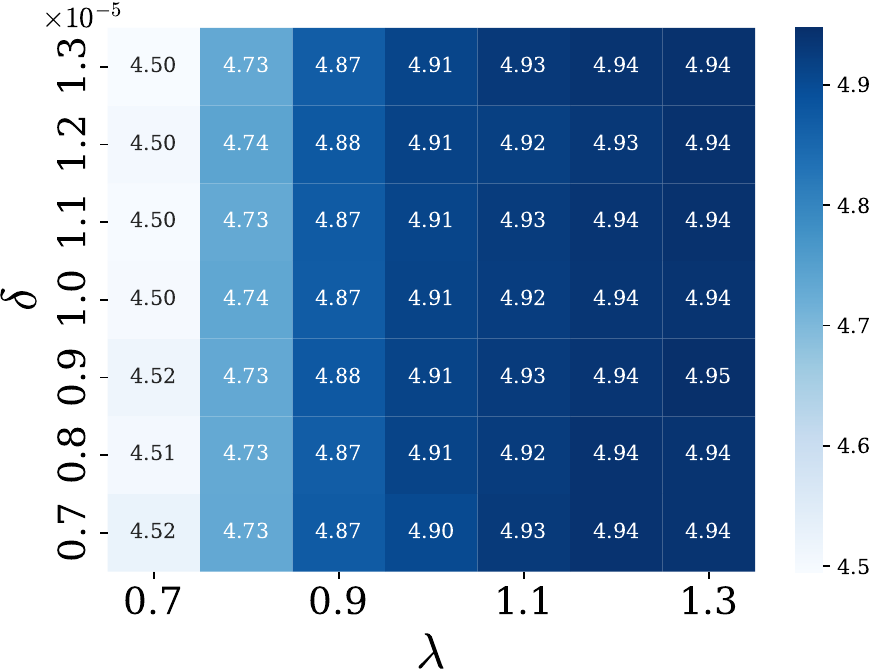}
    \caption{$\lambda$ and $\delta$}
    \label{fig:lambda_delta}
  \end{subfigure}
  \hfill
  \begin{subfigure}{0.24\textwidth}
    \includegraphics[width=\linewidth]{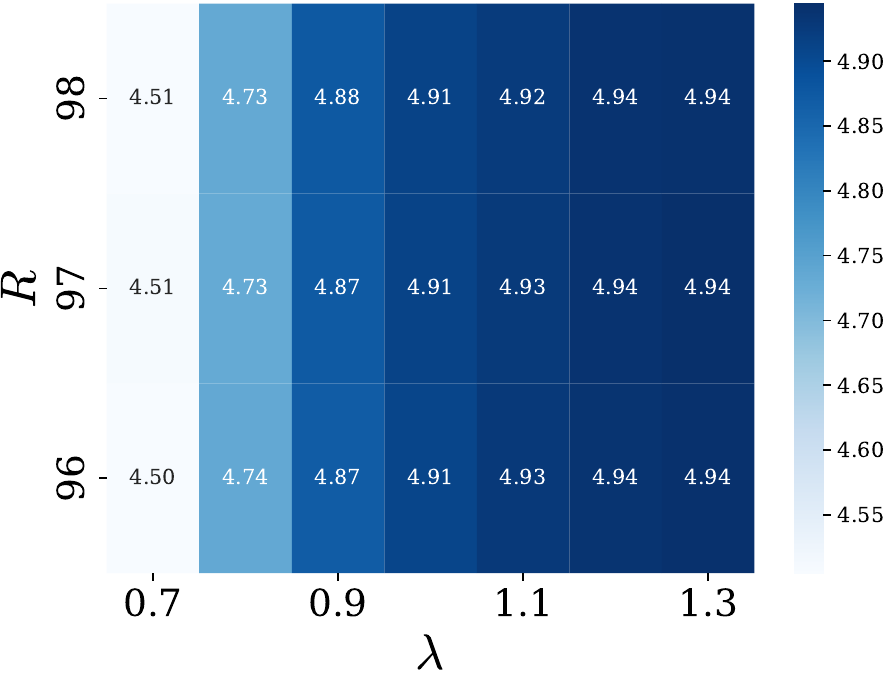}
    \caption{$\lambda$ and $R$}
    \label{fig:lambda_R}
  \end{subfigure}
  \hfill
  \begin{subfigure}{0.24\textwidth}
    \includegraphics[width=\linewidth]{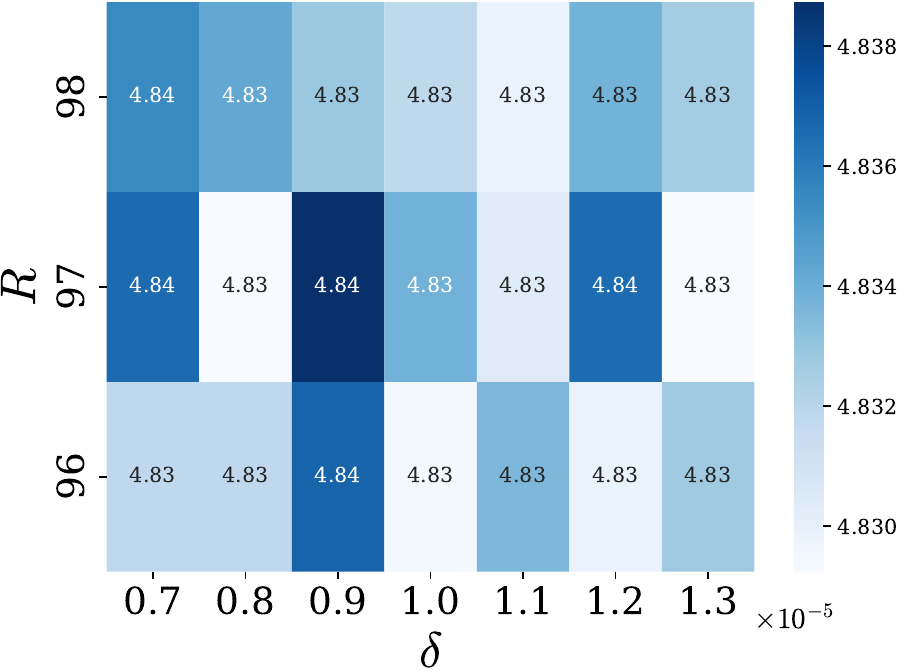}
    \caption{$\delta$ and $R$}
    \label{fig:delta_R}
  \end{subfigure}
  \hfill
  \begin{minipage}{0.24\textwidth}
    \vspace{0pt} 
  \end{minipage}

  \caption{
   Grid search of the required parameters in Diffiner. 
   All scores in the graphs denote NISQA.  
  }
  \label{fig:sigmoid_gridsearch}
\end{figure*}
To decide which hyper-parameters to use in Diffiner, we conducted a grid search based on the following groups:
\begin{description}[labelwidth=3em, labelsep=0.25em]
    \item[$\eta_a$, $\eta_b$] : Updates for DDRM
    \item[$\alpha$, $\beta$, $\gamma$] : For sigmoid-type noise generation in Diffiner for SS
    \item[$\lambda$, $\delta$, $R$] : Clipping of the noise map in Diffiner for SE
\end{description}
As mentioned in the main body, our purpose is to improve the perceptual quality by Diffiner.
Diffiner's outputs contain some generated parts that make reference-based evaluation difficult because they do not correspond to those of the reference signal.
Therefore, we employed a reference-free metrics, i.e., NISQA, as the criterion in this grid search.

The results are summarized in Fig.~\ref{fig:sigmoid_gridsearch}.
Regarding $\eta_a$ and $\eta_b$, we adjusted $\eta_a = 0.9$ and $\eta_b = 0.9$ since they scored the best performance (see Fig.~\ref{fig:sigmoid_gridsearch}(\subref{fig:etas})).
Regarding $\alpha$, $\beta$, and $\gamma$, we employed $\alpha = 2.0$, $\beta = 2.0$, and $\gamma = 0.8$ by monitoring their relationship on NISQA (see Figs.~\ref{fig:sigmoid_gridsearch}(\subref{fig:alpha_beta})--(\subref{fig:alpha_gamma})).
This combination of $\alpha = 2.0$, $\beta = 2.0$, and $\gamma = 0.8$ might not be the best, unlike $\eta_a$ and $\eta_b$, but it is relatively good (as shown in the figures) and the perceptual listening quality obtained by Diffiner when using it was also good.
Regarding $\lambda$, $R$, and $\delta$, we also employed $\lambda = 1.0$, $R = 97$, and $\delta = \num{1.0e-5}$, for the same reason (see Figs.~\ref{fig:sigmoid_gridsearch}(\subref{fig:lambda_delta})--(\subref{fig:delta_R})).\looseness=-1

In addition, we provide a practical guide to select a good set of parameters based on their heatmaps in Fig.~\ref{fig:sigmoid_gridsearch}.
Regarding $\eta_a$ and $\eta_b$, we can see here that as long as $\eta_a=1.0$ is avoided, there is not much variation in performance, indicating relatively robust operation (see Fig.~\ref{fig:sigmoid_gridsearch}(\subref{fig:etas})). 
Regarding $\alpha$, $\beta$, and $\gamma$, there seems to be something like borders of $\alpha = 1.6$ (see Figs.~\ref{fig:sigmoid_gridsearch}(\subref{fig:alpha_beta}) and \ref{fig:sigmoid_gridsearch}(\subref{fig:alpha_beta})), and thus all we have to do is set $\alpha$ to $1.6$ or higher.
Selecting the best set of $\gamma$ and $\beta$ is then relatively easier than setting all of them from scratch.
Regarding $\lambda$, $R$, and $\delta$, $R$ and $\delta$ seem to be insensitive to the performance of NISQA (see the color bar in Fig.~\ref{fig:sigmoid_gridsearch}(\subref{fig:delta_R})).
Then, the colors of ($\lambda$ vs $\delta$) and ($\lambda$ vs $R$) faded gradually from left to right, i.e., along $\lambda$. 
Thus, we can easily choose $\lambda$ first, and next, selecting an arbitrary $R$ and $\delta$ would be fine.

Based on this discussions, it seems that some parameters, i.e., $\alpha$, $R$, and $\delta$, might be relatively insensitive to the performance of NISQA as long as they are set around appropriate values.
Furthermore, the changes in values of any other parameters, i.e., ($\beta$, $\gamma$) and $\lambda$, are generally gradual.
In other words, they have good ranges rather than local values. 
In terms of the model robustness, these characteristics, i.e., insensitiveness of parameters and having good ranges, are practically good for users.
However, these are just an example on our task and dataset, and thus it is preferable to search for the good set of hyper-parameters in accordance with the user's task and dataset.\looseness=-1

\section{Analysis in terms of noise type and gender}
\label{app:noise_type_gender}
To examine the noise robustness of Diffiner in more detail, we compared the performance on real- and unreal-world noise that they have same power degree of noise.
Specifically, we replaced the CHiME noise with either white and pink noise; the target clean speech was left untouched, and we generated white and pink noise with the same root mean square (RMS) as the original CHiME noise to simulate the speech samples mixed with unreal-world noise.

\begin{figure}[t]
 \centering
 \begin{subfigure}{0.24\textwidth}
    \includegraphics[width=\columnwidth]{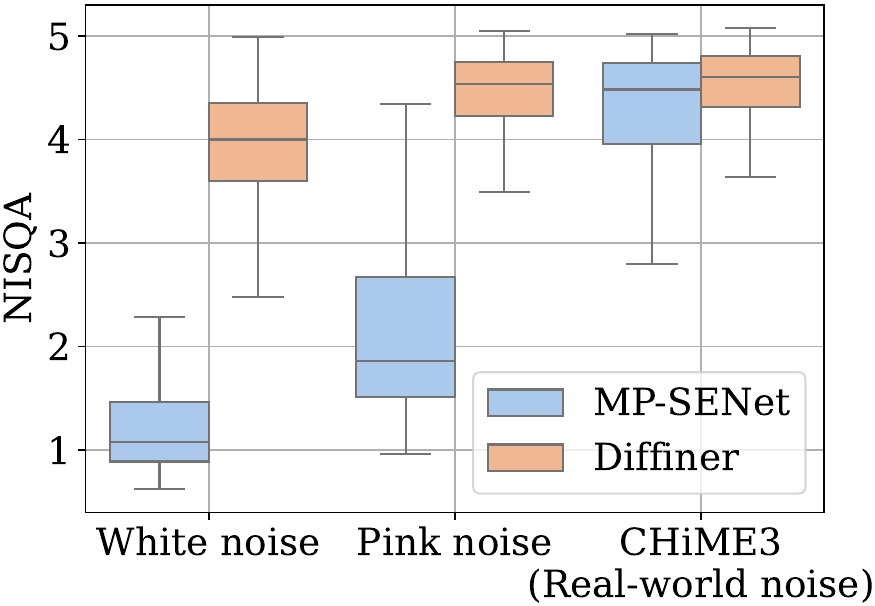}
    \caption{NISQA}
 \end{subfigure}
 \hfill
 \begin{subfigure}{0.24\textwidth}
    \includegraphics[width=\columnwidth]{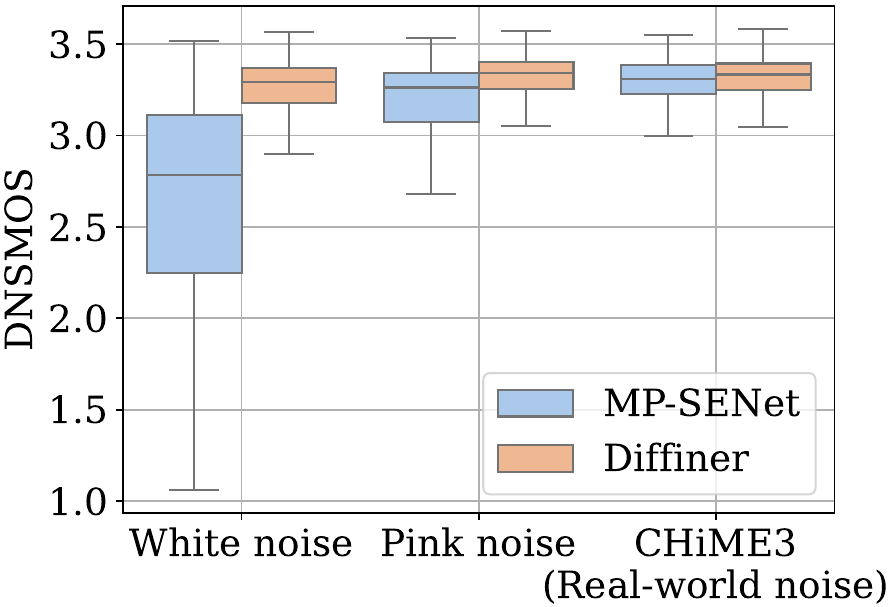}
    \caption{DNSMOS}
 \end{subfigure}  
 \caption{Boxplots comparing the results before and after applying Diffiner across different noise categories.}
 \label{fig:se_compare}
\end{figure}

The performance for NISQA and DNSMOS before and after applying Diffiner is shown in Fig.~\ref{fig:se_compare}.
Note that we used MP-SENet as the preceding model.
For all noise categories, all scores were improved after applying Diffiner compared to the corresponding preceding outputs of MP-SENet.
In particular, Diffiner recovered the scores of unreal-world noise resulting in the same degree of scores as the cases of real-world noise, whereas they were quite degraded when only applying the preceding model (see blue boxes in Figs.~\ref{fig:se_compare}(a) and (b)).
Furthermore, all score variance became comparable or smaller than before applying Diffiner.
These findings imply the robustness and versatility of Diffiner against both real- and unreal-world noise.\looseness=-1

Next, we divided the dataset into three subsets---male-male, male-female, and female-female---and newly conducted performance evaluations before and after applying Diffiner for each subset.
We monitored the performance using reference-based and reference-free metrics, i.e., SI-SDR and NISQA, in these experiments.
Note that we employed Conv-TasNet as the preceding method.\looseness=-1

\begin{figure}[t]
 \centering
 \begin{subfigure}{0.24\textwidth}
    \includegraphics[width=\columnwidth]{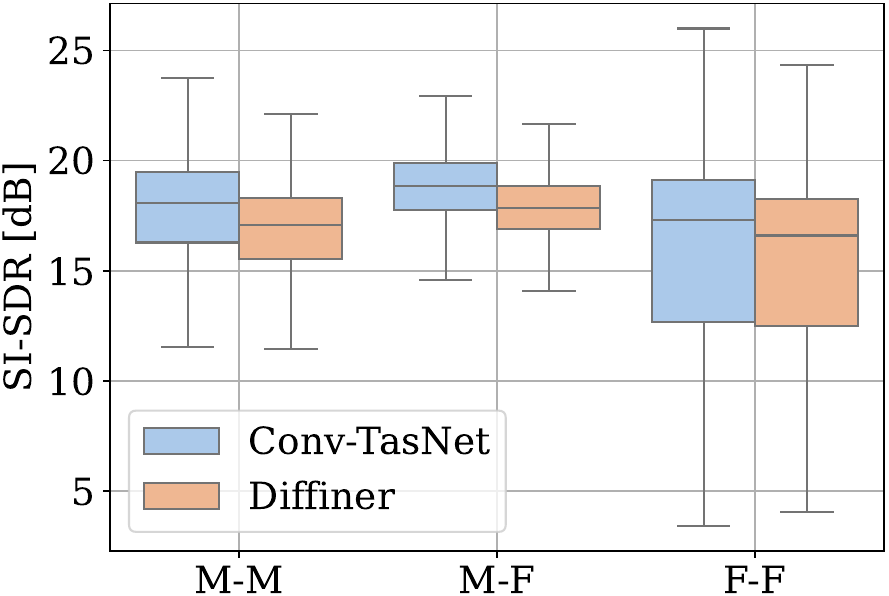}
    \caption{SI-SDR [dB]}
 \end{subfigure}
 \hfill
 \begin{subfigure}{0.24\textwidth}
    \includegraphics[width=\columnwidth]{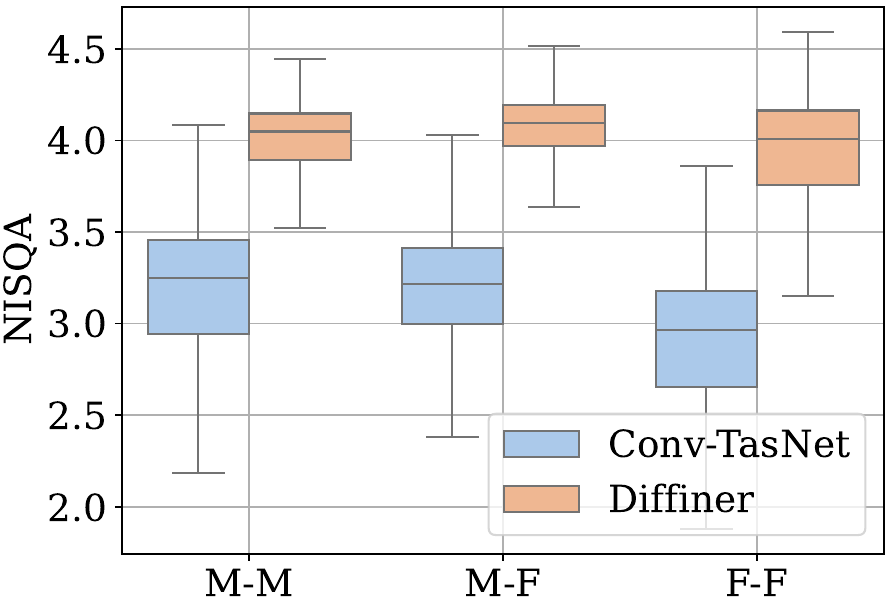}
    \caption{NISQA}
 \end{subfigure}  
 \caption{Boxplots comparing the results before and after applying Diffiner across different gender combinations. 
 Note that `M' = Male speakers, `F' = Female speakers.}
 \label{fig:ss_compare}
\end{figure}

The results are summarized in Fig.~\ref{fig:ss_compare}.
First, regardless of the kind of subset, Diffiner degraded the scores of reference-based metrics while improving those of reference-free metrics.
This aligns with our original experiments (see Table~\ref{table:results_sep}), since Diffiner is a generative model and outputs generated components that are not originally in the corresponding reference.
Furthermore, regardless of before and after applying Diffiner, we can observe that separating speech from the same gender tends to be more difficult than separating speech from different genders.
This suggests that the closer the frequency characteristics of the two speakers, the more challenging the separation becomes, which aligns with intuition. 
However, only in the case of after applying Diffiner, the whisker range became shorter than the corresponding results before applying Diffiner.
This implies that Diffiner would improve not only the perceptual quality but also robustness against gender difference for the task of speech separation.

\section{Phoneme Similarity Before and After Diffiner}
\label{app:phoneme_sim}
\begin{figure}[t]
 \centering
    \includegraphics[scale=0.5]{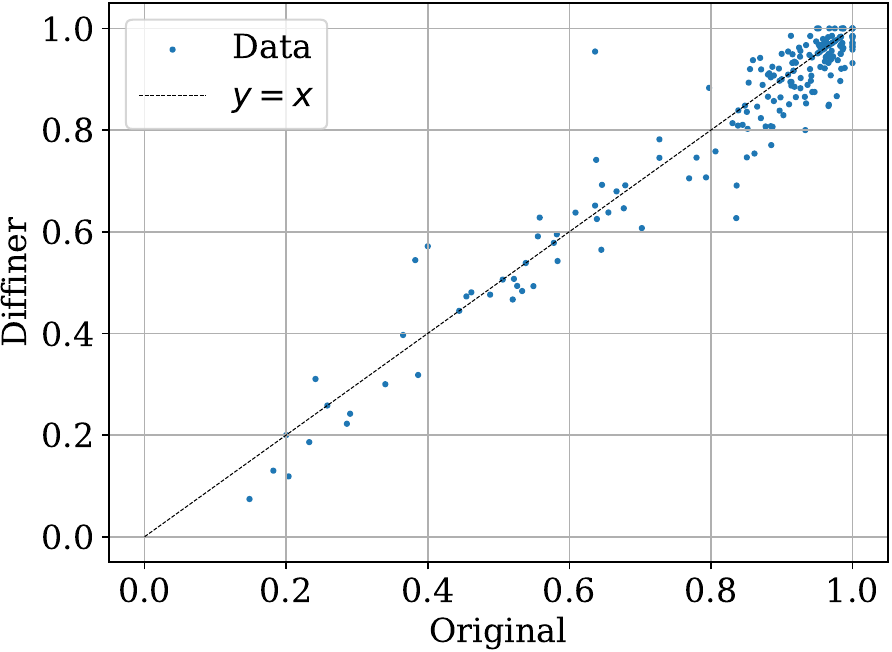}
    \caption{
    Phoneme similarity before and after applying Diffiner to the outputs of Deep Clustering. 
    The horizontal ``Original'' shows the phoneme similarity of Deep Clustering outputs against a clean signal, while the vertical ``Diffiner'' shows that of Diffiner outputs against a clean signal. The correlation coefficient is 0.966.
    }
 \label{fig:phoneme_similarity_rev3}
\end{figure}
To confirm the influence in terms of phoneme similarity, we present a scatter plot of phoneme similarity in Fig.~\ref{fig:phoneme_similarity_rev3}.
We first applied deep clustering (DC) and then put phoneme similarity between DC and the corresponding clean speeches on the $x$-axis, and next applied Diffiner to the DC's output and put phoneme similarity between Diffiner and the corresponding clean speeches on the $y$-axis.
Note that the dashed line represents the line $y=x$ as a criterion; namely, if Diffiner did not spoil the phoneme similarity with clean speech samples, all scatters would ideally be in the upper triangular region.
In the figure, we can see that there is almost no change in phoneme similarity before applying Diffiner ($x$-axis) and after applying Diffiner ($y$-axis). 
This observation is further supported by a quantitative analysis showing that the Pearson correlation coefficient between the two is 0.966, indicating significantly high correlation.

This suggests that Diffiner neither modifies the phonemes contained in the output of the preceding model used for conditioning nor generates clean hallucinated speech.
Furthermore, from the results of our additional experiments on the ASR task (see Fig.~\ref{fig:wer_blend}), Diffiner potentially improves phoneme similarity by using our blending strategy.

While we consider that the phoneme similarity is strongly related to the performance of ASR, our main purpose is refinement of perceptual speech quality.
However, we understand that the refined results would not be adequate if they contained hallucinated mumbled speech, as described in~\cite{eval_generative_se}.
Therefore, in the future, we may consider measuring the temporal speaker consistency of the preceding model output, and then reassigning the correct phonemes in the intervals where the values have degraded, i.e., implying speaker swaps and phoneme errors.

{
\bibliographystyle{IEEEtran}
\bibliography{TASLP_refs}

@inproceedings{lemaguer25_ssw,
  title     = {{Speech Synthesis Evaluation from a voting perspective - a starting point}},
  author    = {S\'{e}bastien {Le Maguer} and Juraj \v{S}imko},
  year      = {2025},
  booktitle = {Proc. of {13th edition of the Speech Synthesis Workshop}},
  pages     = {123--129},
  doi       = {10.21437/SSW.2025-19},
}

@INPROCEEDINGS{eval_generative_se,
  author={Pirklbauer, Jan and Sach, Marvin and Fluyt, Kristoff and Tirry, Wouter and Wardah, Wafaa and Moeller, Sebastian and Fingscheidt, Tim},
  booktitle={Proc. of Speech Communication; 15th ITG Conference}, 
  title={Evaluation Metrics for Generative Speech Enhancement Methods: Issues and Perspectives}, 
  year={2023},
  pages={265-269},
}

@inproceedings{sbctm,
    title={Schr\"odinger Bridge Consistency Trajectory Models for Speech Enhancement}, 
    author={Shuichiro Nishigori and Koichi Saito and Naoki Murata and Masato Hirano and Shusuke Takahashi and Yuki Mitsufuji},
    booktitle = {Proc. of IEEE Workshop on Applications of Signal Processing to Audio and Acoustics (WASPAA)},
    year={2025},
}

@inproceedings{sb_se,
  title     = {{Schr\"odinger Bridge for Generative Speech Enhancement}},
  author    = {Ante Juki\'c and Roman Korostik and Jagadeesh Balam and Boris Ginsburg},
  year      = {2024},
  booktitle = {Proc. of {Interspeech}},
  pages     = {1175--1179},
}

@inproceedings{schrodinger_bridge,
    title={Likelihood Training of Schr\"odinger Bridge using Forward-Backward {SDE}s Theory}, 
    author={Tianrong Chen and Guan-Horng Liu and Evangelos A. Theodorou},
    booktitle={Proc. of International Conference on Learning Representations (ICLR)},
    year={2021},
}

@inproceedings{soundctm,
  title={Sound{CTM}: Unifying Score-based and Consistency Models for Full-band Text-to-Sound Generation},
  author={Koichi Saito and Dongjun Kim and Takashi Shibuya and Chieh-Hsin Lai and Zhi Zhong and Yuhta Takida and Yuki Mitsufuji},
  booktitle={Proc. of International Conference on Learning Representations (ICLR)},
  year={2025},
}

@inproceedings{ctm,
  title={Consistency Trajectory Models: Learning Probability Flow {ODE} Trajectory of Diffusion},
  author={Dongjun Kim and Chieh-Hsin Lai and Wei-Hsiang Liao and Naoki Murata and Yuhta Takida and Toshimitsu Uesaka and Yutong He and Yuki Mitsufuji and Stefano Ermon},
  booktitle={Proc. of International Conference on Learning Representations (ICLR)},
  year={2024},
}

@inproceedings{voicebankdemand,
  author={Cassia Valentini-Botinhao and Xin Wang and Shinji Takaki and Junichi Yamagishi},
  title={Investigating {RNN}-based speech enhancement methods for noise-robust Text-to-Speech},
  year=2016,
  booktitle={Proc. of International Speech Communication Association (ISCA) Speech Synthesis Workshop},
  pages={146--152}
}

@misc{ljspeech,
  author       = {Keith Ito and Linda Johnson},
  title        = {The LJ Speech Dataset},
  howpublished = {\url{https://keithito.com/LJ-Speech-Dataset/}},
  year         = 2017
}

@ARTICLE{se_datasize,
  author={Gonzalez, Philippe and Tan, Zheng-Hua and Østergaard, Jan and Jensen, Jesper and Alstrøm, Tommy Sonne and May, Tobias},
  journal={IEEE Signal Processing Letters}, 
  title={The Effect of Training Dataset Size on Discriminative and Diffusion-Based Speech Enhancement Systems}, 
  year={2024},
  volume={31},
  number={},
  pages={2225-2229},
  doi={10.1109/LSP.2024.3449221}
}

@INPROCEEDINGS{pit_org,
  author={Yu, Dong and Kolbæk, Morten and Tan, Zheng-Hua and Jensen, Jesper},
  booktitle={Proc. of IEEE International Conference on Acoustics, Speech and Signal Processing (ICASSP)}, 
  title={Permutation invariant training of deep models for speaker-independent multi-talker speech separation}, 
  year={2017},
  volume={},
  number={},
  pages={241-245},
  doi={10.1109/ICASSP.2017.7952154}
}

@INPROCEEDINGS{rnn_se_3,
  author={Xia, Yangyang and Braun, Sebastian and Reddy, Chandan K. A. and Dubey, Harishchandra and Cutler, Ross and Tashev, Ivan},
  booktitle={Proc. of IEEE International Conference on Acoustics, Speech and Signal Processing (ICASSP)},
  title={Weighted Speech Distortion Losses for Neural-Network-Based Real-Time Speech Enhancement},
  year={2020},
  pages={871-875},
}

@inproceedings{rnn_se_2,
  author={Nils L. Westhausen and Bernd T. Meyer},
  title={{Dual-Signal Transformation LSTM Network for Real-Time Noise Suppression}},
  year=2020,
  booktitle={Proc. of Interspeech},
  pages={2477--2481},
}

@INPROCEEDINGS{crnn_hybrid_se_3,
  author={Choi, Hyeong-Seok and Park, Sungjin and Lee, Jie Hwan and Heo, Hoon and Jeon, Dongsuk and Lee, Kyogu},
  booktitle={Proc. of IEEE International Conference on Acoustics, Speech and Signal Processing (ICASSP)},
  title={Real-Time Denoising and Dereverberation with Tiny Recurrent {U}-Net},
  year={2021},
  pages={5789-5793},
}

@INPROCEEDINGS{rnn_se_1,
  author={Hao, Xiang and Su, Xiangdong and Horaud, Radu and Li, Xiaofei},
  booktitle={Proc. of IEEE International Conference on Acoustics, Speech and Signal Processing (ICASSP)},
  title={Fullsubnet: A Full-Band and Sub-Band Fusion Model for Real-Time Single-Channel Speech Enhancement},
  year={2021},
  pages={6633-6637},
}

@inproceedings{nmf_org,
 author = {Lee, Daniel and Seung, H. Sebastian},
 booktitle = {Advances in Neural Information Processing Systems},
 pages = {},
 title = {Algorithms for Non-negative Matrix Factorization},
 volume = {13},
 year = {2000}
}

@ARTICLE{subtraction_org,
  author={Boll, S.},
  journal={IEEE Transactions on Acoustics, Speech, and Signal Processing}, 
  title={Suppression of acoustic noise in speech using spectral subtraction}, 
  year={1979},
  volume={27},
  number={2},
  pages={113-120},
  doi={10.1109/TASSP.1979.1163209}
}

@book{wiener_org,
    author = {Wiener, Norbert},
    title = "{Extrapolation, Interpolation, and Smoothing of Stationary Time Series: With Engineering Applications}",
    publisher = {The MIT Press},
    year = {1949},
    month = {08},
    isbn = {9780262257190},
    doi = {10.7551/mitpress/2946.001.0001},
}

@InProceedings{se_traditional_survey,
author="Gaddamedi, Satya Prasad
and Patel, Anuj
and Chandra, Sabyasachi
and Bharati, Puja
and Ghosh, Nirmalya
and Mandal, Shyamal Kumar Das",
title="Speech Enhancement: Traditional and Deep Learning Techniques",
booktitle="Proceedings of 27th International Symposium on Frontiers of Research in Speech and Music",
year="2024",
pages="75--86",
isbn="978-981-97-1549-7"
}

@article{ss_survey,
author = {Mirbeygi, Mohaddeseh and Mahabadi, Aminollah and Ranjbar, Akbar},
title = {Speech and music separation approaches - a survey},
year = {2022},
issue_date = {Jun 2022},
publisher = {Kluwer Academic Publishers},
address = {USA},
volume = {81},
number = {15},
issn = {1380-7501},
doi = {10.1007/s11042-022-11994-1},
journal = {Multimedia Tools Appl.},
pages = {21155–21197},
numpages = {43}
}

@ARTICLE{se_survey,
  author={O'Shaughnessy, Douglas},
  journal={IEEE Transactions on Human-Machine Systems}, 
  title={Speech Enhancement—A Review of Modern Methods}, 
  year={2024},
  volume={54},
  number={1},
  pages={110-120},
  doi={10.1109/THMS.2023.3339663}
}

@article{speaker_diar_survey,
title = {A review of speaker diarization: Recent advances with deep learning},
journal = {Computer Speech \& Language},
volume = {72},
pages = {101317},
year = {2022},
issn = {0885-2308},
doi = {https://doi.org/10.1016/j.csl.2021.101317},
author = {Tae Jin Park and Naoyuki Kanda and Dimitrios Dimitriadis and Kyu J. Han and Shinji Watanabe and Shrikanth Narayanan},
}

@ARTICLE{speaker_id_survey,
  author={Kabir, Muhammad Mohsin and Mridha, M. F. and Shin, Jungpil and Jahan, Israt and Ohi, Abu Quwsar},
  journal={IEEE Access}, 
  title={A Survey of Speaker Recognition: Fundamental Theories, Recognition Methods and Opportunities}, 
  year={2021},
  volume={9},
  number={},
  pages={79236-79263},
  doi={10.1109/ACCESS.2021.3084299}
}

@ARTICLE{asr_survey,
  author={Prabhavalkar, Rohit and Hori, Takaaki and Sainath, Tara N. and Schlüter, Ralf and Watanabe, Shinji},
  journal={IEEE/ACM Transactions on Audio, Speech, and Language Processing}, 
  title={{End-to-End} Speech Recognition: A Survey}, 
  year={2024},
  volume={32},
  number={},
  pages={325-351},
  doi={10.1109/TASLP.2023.3328283}
}

@article{se_vocoder_1,
  author = {Shi, Jing and Chang, Xuankai and Hayashi, Tomoki and Lu, Yen-Ju and Watanabe, Shinji and Xu, Bo},
  title = {Discretization and Re-synthesis: an alternative method to solve the Cocktail Party Problem},
  journal   = {arXiv preprint arXiv:2112.09382},
  year = {2021},
}

@inproceedings{hershey2016deep,
  title={Deep clustering: Discriminative embeddings for segmentation and separation},
  author={Hershey, John R and Chen, Zhuo and Le Roux, Jonathan and Watanabe, Shinji},
  booktitle={IEEE International Conference on Acoustics, Speech and Signal Processing (ICASSP)},
  pages={31--35},
  year={2016}
}

@article{luo2019conv,
  title={Conv-tasnet: Surpassing ideal time--frequency magnitude masking for speech separation},
  author={Luo, Yi and Mesgarani, Nima},
  journal={IEEE/ACM Transactions on Audio, Speech, and Language Processing (TASLP)},
  volume={27},
  number={8},
  pages={1256--1266},
  year={2019},
  publisher={IEEE}
}

@inproceedings{subakan2021attention,
  title={Attention is all you need in speech separation},
  author={Subakan, Cem and Ravanelli, Mirco and Cornell, Samuele and Bronzi, Mirko and Zhong, Jianyuan},
  booktitle={IEEE International Conference on Acoustics, Speech and Signal Processing (ICASSP)},
  pages={21--25},
  year={2021}
}

@inproceedings{rix2001perceptual,
  title={Perceptual evaluation of speech quality ({PESQ})-a new method for speech quality assessment of telephone networks and codecs},
  author={Rix, Antony W and Beerends, John G and Hollier, Michael P and Hekstra, Andries P},
  booktitle={IEEE International Conference on Acoustics, Speech and Signal Processing (ICASSP)},
  volume={2},
  pages={749--752},
  year={2001}
}

@inproceedings{taal2010short,
  title={A short-time objective intelligibility measure for time-frequency weighted noisy speech},
  author={Taal, Cees H and Hendriks, Richard C and Heusdens, Richard and Jensen, Jesper},
  booktitle={Proc. of IEEE international conference on acoustics, speech and signal processing (ICASSP)},
  pages={4214--4217},
  year={2010},
  organization={IEEE}
}

@article{kingma2013auto,
  title={Auto-encoding variational bayes},
  author={Kingma, Diederik P and Welling, Max},
  journal={arXiv preprint arXiv:1312.6114},
  year={2013}
}

@article{creswell2018generative,
  title={Generative adversarial networks: An overview},
  author={Creswell, Antonia and White, Tom and Dumoulin, Vincent and Arulkumaran, Kai and Sengupta, Biswa and Bharath, Anil A},
  journal={IEEE signal processing magazine},
  volume={35},
  number={1},
  pages={53--65},
  year={2018},
  publisher={IEEE}
}

@article{ho2020denoising,
  title={Denoising diffusion probabilistic models},
  author={Ho, Jonathan and Jain, Ajay and Abbeel, Pieter},
  journal={Advances in Neural Information Processing Systems},
  volume={33},
  pages={6840--6851},
  year={2020}
}

@article{kong2020diffwave,
  title={Diffwave: A versatile diffusion model for audio synthesis},
  author={Kong, Zhifeng and Ping, Wei and Huang, Jiaji and Zhao, Kexin and Catanzaro, Bryan},
  journal={arXiv preprint arXiv:2009.09761},
  year={2020}
}

@article{richter2023speech,
  title={Speech enhancement and dereverberation with diffusion-based generative models},
  author={Richter, Julius and Welker, Simon and Lemercier, Jean-Marie and Lay, Bunlong and Gerkmann, Timo},
  journal={IEEE/ACM Transactions on Audio, Speech, and Language Processing (TASLP)},
  year={2023},
  publisher={IEEE}
}

@article{lemercier2022storm,
  author={Lemercier, Jean-Marie and Richter, Julius and Welker, Simon and Gerkmann, Timo},
  journal={IEEE/ACM Transactions on Audio, Speech, and Language Processing (TASLP)}, 
  title={{StoRM: A Diffusion-Based Stochastic Regeneration Model for Speech Enhancement and Dereverberation}}, 
  year={2023},
  volume={31},
  number={},
  pages={2724-2737},
  doi={10.1109/TASLP.2023.3294692}}

@inproceedings{sawata23_interspeech,
  author={Ryosuke Sawata and Naoki Murata and Yuhta Takida and Toshimitsu Uesaka and Takashi Shibuya and Shusuke Takahashi and Yuki Mitsufuji},
  title={{Diffiner: A Versatile Diffusion-based Generative Refiner for Speech Enhancement}},
  year=2023,
  booktitle={Proc. INTERSPEECH},
  pages={3824--3828},
  doi={10.21437/Interspeech.2023-1547},
  issn={2958-1796}
}

@inproceedings{jayaram2020source,
  title={Source separation with deep generative priors},
  author={Jayaram, Vivek and Thickstun, John},
  booktitle={International Conference on Machine Learning (ICML)},
  pages={4724--4735},
  year={2020}
}

@inproceedings{scheibler2022diffusion,
  author={Scheibler, Robin and Ji, Youna and Chung, Soo-Whan and Byun, Jaeuk and Choe, Soyeon and Choi, Min-Seok},
  booktitle={Proc. of IEEE International Conference on Acoustics, Speech and Signal Processing (ICASSP)}, 
  title={Diffusion-Based Generative Speech Source Separation}, 
  year={2023},
  volume={},
  number={},
  pages={1-5},
  doi={10.1109/ICASSP49357.2023.10095310}
}

@article{lutati2023separate,
  title={Separate And Diffuse: Using a Pretrained Diffusion Model for Improving Source Separation},
  author={Lutati, Shahar and Nachmani, Eliya and Wolf, Lior},
  journal={arXiv preprint arXiv:2301.10752},
  year={2023}
}

@inproceedings{kawar2022denoising,
    title={Denoising Diffusion Restoration Models},
    author={Bahjat Kawar and Michael Elad and Stefano Ermon and Jiaming Song},
    booktitle={Advances in Neural Information Processing Systems},
    year={2022}
}

@inproceedings{mittag2021nisqa,
  author={Gabriel Mittag and Babak Naderi and Assmaa Chehadi and Sebastian Möller},
  title={{NISQA: A Deep CNN-Self-Attention Model for Multidimensional Speech Quality Prediction with Crowdsourced Datasets}},
  year=2021,
  booktitle={Proc. INTERSPEECH},
  pages={2127--2131},
  doi={10.21437/Interspeech.2021-299},
  issn={2958-1796}
}

@article{dhariwal2021diffusion,
  title={Diffusion models beat gans on image synthesis},
  author={Dhariwal, Prafulla and Nichol, Alexander},
  journal={Advances in Neural Information Processing Systems},
  volume={34},
  pages={8780--8794},
  year={2021}
}

@InProceedings{kingma2014adam,
  author    = {Kingma, Diederik and Ba, Jimmy},
  booktitle = {International Conference on Learning Representations (ICLR)},
  title     = {Adam: A Method for Stochastic Optimization},
  year      = {2015},
  optmonth  = {12},
}

@article{song2020improved,
  title={Improved techniques for training score-based generative models},
  author={Song, Yang and Ermon, Stefano},
  journal={Advances in Neural Information Processing Systems},
  volume={33},
  pages={12438--12448},
  year={2020}
}

@article{jensen2016algorithm,
  title={An algorithm for predicting the intelligibility of speech masked by modulated noise maskers},
  author={Jensen, Jesper and Taal, Cees H},
  journal={IEEE/ACM Transactions on Audio, Speech, and Language Processing (TASLP)},
  volume={24},
  number={11},
  pages={2009--2022},
  year={2016},
  publisher={IEEE}
}

@inproceedings{le2019sdr,
  title={{SDR}--half-baked or well done?},
  author={Le Roux, Jonathan and Wisdom, Scott and Erdogan, Hakan and Hershey, John R},
  booktitle={IEEE International Conference on Acoustics, Speech and Signal Processing (ICASSP)},
  pages={626--630},
  year={2019}
}

@article{liu2022voicefixer,
  title={VoiceFixer: A Unified Framework for High-Fidelity Speech Restoration},
  author={Liu, Haohe and Liu, Xubo and Kong, Qiuqiang and Tian, Qiao and Zhao, Yan and Wang, DeLiang and Huang, Chuanzeng and Wang, Yuxuan},
  journal={arXiv preprint arXiv:2204.05841},
  year={2022}
}

@inproceedings{luo2020dual,
  title={Dual-path {RNN}: efficient long sequence modeling for time-domain single-channel speech separation},
  author={Luo, Yi and Chen, Zhuo and Yoshioka, Takuya},
  booktitle={IEEE International Conference on Acoustics, Speech and Signal Processing (ICASSP)},
  pages={46--50},
  year={2020}
}

@inproceedings{reddy2022dnsmos,
  title={{DNSMOS} P. 835: A non-intrusive perceptual objective speech quality metric to evaluate noise suppressors},
  author={Reddy, Chandan KA and Gopal, Vishak and Cutler, Ross},
  booktitle={IEEE International Conference on Acoustics, Speech and Signal Processing (ICASSP)},
  pages={886--890},
  year={2022}
}

@inproceedings{stoller2018wave,
  title={Wave-u-net: A multi-scale neural network for end-to-end audio source separation},
  author={Stoller, Daniel and Ewert, Sebastian and Dixon, Simon},
  booktitle={International Society for Music Information Retrieval (ISMIR)},
  year={2018},
  pages={334-340},
}

@inproceedings{choi2018phase,
  title={Phase-aware speech enhancement with deep complex u-net},
  author={Choi, Hyeong-Seok and Kim, Jang-Hyun and Huh, Jaesung and Kim, Adrian and Ha, Jung-Woo and Lee, Kyogu},
  booktitle={International Conference on Learning Representations (ICLR)},
  year={2018}
}

@inproceedings{pascual2017segan,
  title={SEGAN: Speech enhancement generative adversarial network},
  author={Pascual, Santiago and Bonafonte, Antonio and Serra, Joan},
  booktitle={Proc. INTERSPEECH},
  pages={3642-3646},
  year={2017}
}

@article{serra2022universal,
  title={Universal speech enhancement with score-based diffusion},
  author={Serr{\`a}, Joan and Pascual, Santiago and Pons, Jordi and Araz, R Oguz and Scaini, Davide},
  journal={arXiv preprint arXiv:2206.03065},
  year={2022}
}

@article{lu2023explicit,
  title={Explicit Estimation of Magnitude and Phase Spectra in Parallel for High-Quality Speech Enhancement},
  author={Lu, Ye-Xin and Ai, Yang and Ling, Zhen-Hua},
  journal={arXiv preprint arXiv:2308.08926},
  year={2023}
}

@inproceedings{zhao2021monaural,
  title={Monaural speech enhancement with complex convolutional block attention module and joint time frequency losses},
  author={Zhao, Shengkui and Nguyen, Trung Hieu and Ma, Bin},
  booktitle={IEEE International Conference on Acoustics, Speech and Signal Processing (ICASSP)},
  pages={6648--6652},
  year={2021}
}

@inproceedings{lu2023mp,
  author={Ye-Xin Lu and Yang Ai and Zhen-Hua Ling},
  title={{MP-SENet: A Speech Enhancement Model with Parallel Denoising of Magnitude and Phase Spectra}},
  year=2023,
  booktitle={Proc. INTERSPEECH},
  pages={3834--3838},
  doi={10.21437/Interspeech.2023-1441},
  issn={2958-1796}
}

@inproceedings{barker2015third,
  title={The third ‘{CHiME}’ speech separation and recognition challenge: Dataset, task and baselines},
  author={Barker, Jon and Marxer, Ricard and Vincent, Emmanuel and Watanabe, Shinji},
  booktitle={IEEE Workshop on Automatic Speech Recognition and Understanding (ASRU)},
  pages={504--511},
  year={2015}
}

@article{hu2020dccrn,
  title={DCCRN: Deep complex convolution recurrent network for phase-aware speech enhancement},
  author={Hu, Yanxin and Liu, Yun and Lv, Shubo and Xing, Mengtao and Zhang, Shimin and Fu, Yihui and Wu, Jian and Zhang, Bihong and Xie, Lei},
  journal={Proc. INTERSPEECH},
  year={2020},
  pages={2472-2476}
}

@inproceedings{li2024diffusion,
  title={Diffusion-based generative modeling with discriminative guidance for streamable speech enhancement},
  author={Li, Chenda and Cornell, Samuele and Watanabe, Shinji and Qian, Yanmin},
  booktitle={Proc. of IEEE Spoken Language Technology Workshop (SLT)},
  pages={333--340},
  year={2024},
  organization={IEEE}
}

@article{bureau2001method,
  title={Method for the subjective assessment of intermediate quality level of coding systems},
  author={Bureau, ITU Radiocommunication},
  journal={Recommendation ITU-R BS. 1534},
  year={2001}
}

@inproceedings{radford2023robust,
  title={Robust speech recognition via large-scale weak supervision},
  author={Radford, Alec and Kim, Jong Wook and Xu, Tao and Brockman, Greg and McLeavey, Christine and Sutskever, Ilya},
  booktitle={International conference on machine learning},
  pages={28492--28518},
  year={2023},
  organization={PMLR}
}

@article{machavcek2023turning,
  title={Turning whisper into real-time transcription system},
  author={Mach{\'a}{\v{c}}ek, Dominik and Dabre, Raj and Bojar, Ond{\v{r}}ej},
  journal={arXiv preprint arXiv:2307.14743},
  year={2023}
}

@article{baevski2020wav2vec,
  title={wav2vec 2.0: A framework for self-supervised learning of speech representations},
  author={Baevski, Alexei and Zhou, Yuhao and Mohamed, Abdelrahman and Auli, Michael},
  journal={Advances in neural information processing systems},
  volume={33},
  pages={12449--12460},
  year={2020}
}

@article{lee1990overview,
  title={An overview of the SPHINX speech recognition system},
  author={Lee, K-F and Hon, H-W and Reddy, Raj},
  journal={IEEE Transactions on Acoustics, Speech, and Signal Processing},
  volume={38},
  number={1},
  pages={35--45},
  year={1990},
  publisher={IEEE}
}
}

\begin{IEEEbiography}[{\includegraphics[width=1in,height=1.25in,clip,keepaspectratio]{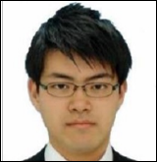}}]{Masato Hirano} received B.S. and M.S. degrees in Applied Physics and Physico-Informatics from Keio University, Japan in 2017 and 2019, respectively.
He is currently a researcher at Sony Group Corporation, Japan.
His research interests include acoustic signal processing and machine learning.\looseness=-1
\end{IEEEbiography}
\begin{IEEEbiography}[{\includegraphics[width=1in,height=1.25in,clip,keepaspectratio]{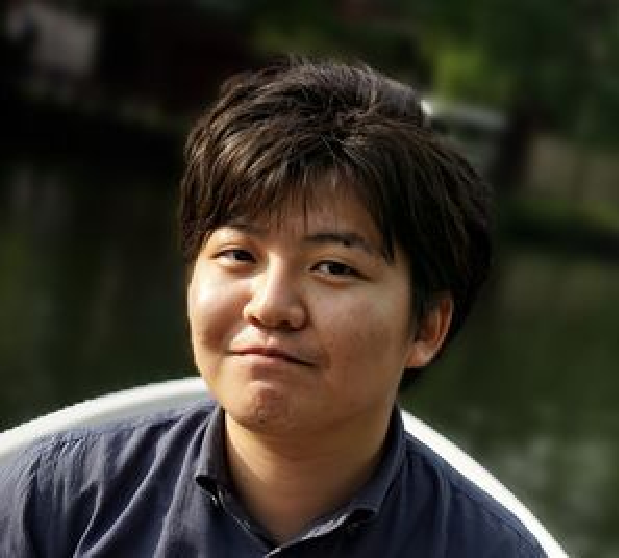}}]{Ryosuke Sawata} received B.S. and M.S. degrees in Electronics and Information Engineering from Hokkaido University, Japan in 2014 and 2016, respectively.
He formerly worked at the Stanford Vision and Learning Lab (SVL) at Stanford University, USA.
He is currently a researcher at Sony Research and a Ph.D. candidate at the Graduate School of Information Science and Technology, Hokkaido University.
His research interests include biosignal processing, acoustic signal processing, and 3D computer vision. 
He is a member of IEEE.\looseness=-1
\end{IEEEbiography}
\begin{IEEEbiography}[{\includegraphics[width=1in,height=1.25in,clip,keepaspectratio]{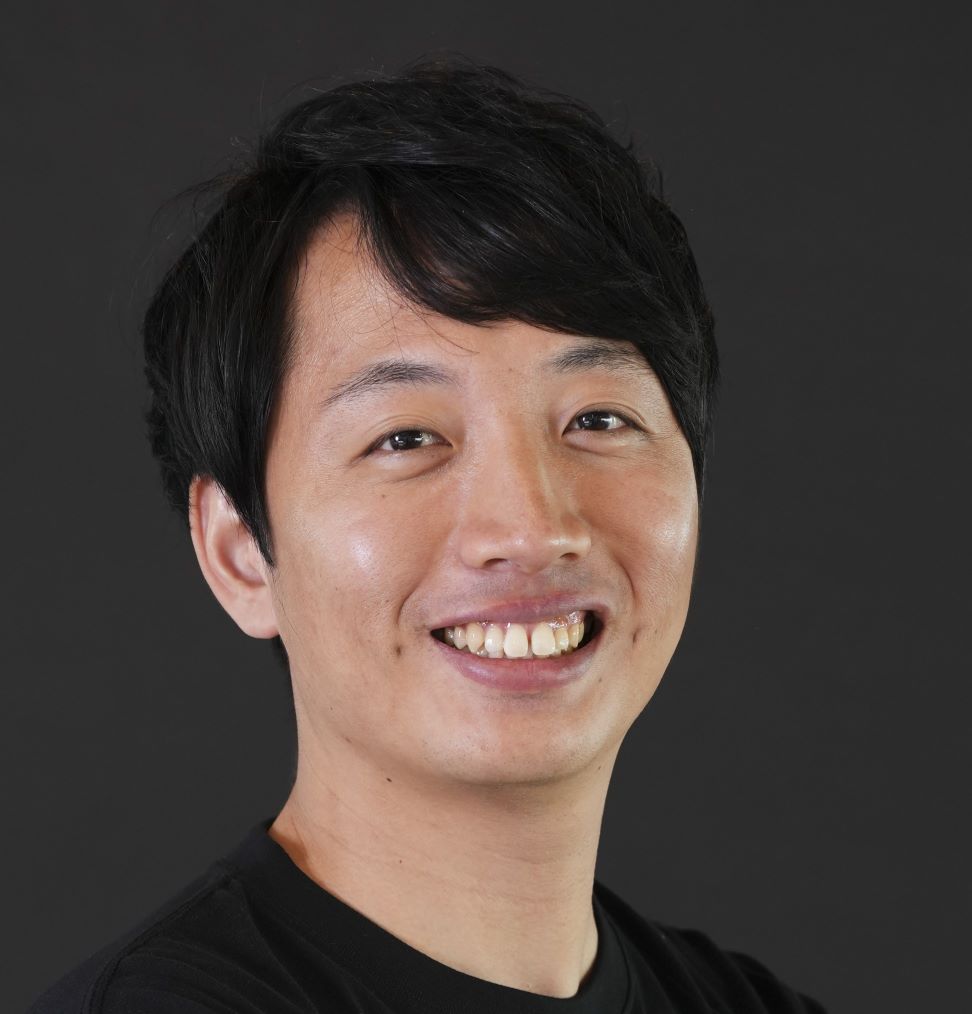}}]{Naoki Murata} received the B.E. degree in engineering and the M.S. degree in information science and technology from the University of Tokyo, Japan, in 2015 and 2017, respectively. He is currently a researcher at Sony Research, Japan. His research interests include deep generative modeling, signal processing, and numerical optimization.\looseness=-1
\end{IEEEbiography}
\begin{IEEEbiography}[{\includegraphics[width=1in,height=1.25in,clip,keepaspectratio]{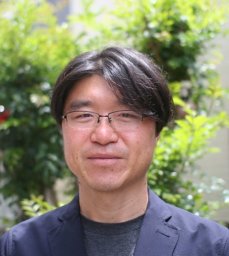}}]{Shusuke Takahashi}
received his B.E. degree in communications engineering and M.S. degree in information science from Tohoku University, Japan, in 2000 and 2002, respectively. He is currently a researcher and a deputy general manager in the Creative AI Lab at Sony Group Corporation, Japan. His research interests include speech and audio signal processing and machine learning. His team achieved the first place in DCASE 2021 Challenge in Task3, and co-organized DCASE Challenge in 2022 and 2023. He also co-organized Sound Demixing Challenge 2023, where real audio from movies by Sony Pictures Entertainment were used for the evaluation.\looseness=-1
\end{IEEEbiography}
\begin{IEEEbiography}[{\includegraphics[width=1in,height=1.25in,clip,keepaspectratio]{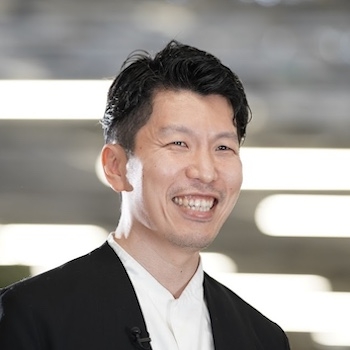}}]{Yuki Mitsufuji} received the B.S. and M.S. degrees in information science from Keio University in 2002 and 2004, respectively, and the Ph.D. degree in information science and technology from the University of Tokyo in 2020.
He leads the Creative AI Lab at Sony Group Corporation. He has been a Visiting Research Professor at New York University since 2025. He previously served as a Specially Appointed Associate Professor at the Tokyo Institute of Technology from 2022 to 2025. He has been selected for the Stanford/Elsevier World's Top 2\% Scientists list.
He joined Sony Corporation in 2004, where he led teams developing the sound design for the PlayStation title Gran Turismo Sport and the spatial audio system Sonic Surf VR. His work has received several honors, including the TIGA Award for Best Audio Design for Gran Turismo Sport and a Jury Selection at the Japan Media Arts Festival for the 576-channel sound field synthesis project Acoustic Vessel Odyssey.
From 2011 to 2012, he was a Visiting Researcher at the Analysis/Synthesis Team at IRCAM in Paris, France. In 2021, his team organized the Music Demixing (MDX) Challenge on AIcrowd using a professionally produced dataset provided by Sony Music, and achieved first place in Task 3 of the DCASE2021 Challenge.\looseness=-1
\end{IEEEbiography}

 




\vfill

\end{document}